\documentclass[11pt]{article}
\pdfoutput=1
\usepackage{fullpage}
\usepackage{cite}
\usepackage{graphicx}
\usepackage{amsmath}
\usepackage{amssymb}
\title{}
\date{}

\def\beq{\begin{equation}}
\def\eeq{\end{equation}}
\allowdisplaybreaks
\begin{document}
\bibliographystyle{utphys}
\newcommand{\msbar}{\ensuremath{\overline{\text{MS}}}}
\newcommand{\DIS}{\ensuremath{\text{DIS}}}
\newcommand{\abar}{\ensuremath{\bar{\alpha}_S}}
\newcommand{\bb}{\ensuremath{\bar{\beta}_0}}
\newcommand{\rc}{\ensuremath{r_{\text{cut}}}}
\newcommand{\Nd}{\ensuremath{N_{\text{d.o.f.}}}}
\setlength{\parindent}{0pt}

\titlepage
\begin{flushright}
BOW-PH-164  \\
QMUL-PH-16-19
\end{flushright}

\vspace*{0.5cm}

\begin{center}
{\bf \Large Next-to-soft corrections to high energy scattering \\
in QCD and gravity}

\vspace*{1cm}
\textsc{A. Luna$^a$\footnote{a.luna-godoy.1@research.gla.ac.uk},
  S. Melville$^a$\footnote{s.melville.1@research.gla.ac.uk},
  S. G. Naculich$^b$\footnote{naculich@bowdoin.edu}, and
  C. D. White$^c$\footnote{Christopher.White@qmul.ac.uk} } \\

\vspace*{0.5cm} $^a$ SUPA, School of Physics and Astronomy, University of Glasgow,\\ Glasgow G12 8QQ, Scotland, UK\\

\vspace*{0.5cm} $^b$ Department of Physics, Bowdoin College, Brunswick, ME 04011, USA\\

\vspace*{0.5cm} $^c$ Centre for Research in String Theory, School of Physics and Astronomy, \\
Queen Mary University of London, 327 Mile End Road, London E1 4NS, UK \\

\end{center}

\vspace*{0.5cm}

\begin{abstract}
We examine the Regge (high energy) limit of 4-point scattering in both
QCD and gravity, using recently developed techniques to systematically
compute all corrections up to next-to-leading power in the exchanged
momentum i.e. beyond the eikonal approximation. We consider the
situation of two scalar particles of arbitrary mass, thus generalising
previous calculations in the literature. In QCD, our calculation
describes power-suppressed corrections to the Reggeisation of the
gluon. In gravity, we confirm a previous conjecture that next-to-soft
corrections correspond to two independent deflection angles for the
incoming particles. Our calculations in QCD and gravity are consistent
with the well-known double copy relating amplitudes in the two
theories.
\end{abstract}

\vspace*{0.5cm}

\section{Introduction}
Scattering amplitudes have many theoretical and phenomenological
applications in (non-)abelian gauge theories and gravity, whilst also
revealing how different theories are related. When studying
amplitudes, it can be useful to consider particular kinematic limits
of scattering processes, which allow all-order insights into the
structure of perturbative quantum field theory. One such limit is the
{\it Regge limit}, in which the centre of mass energy of the
scattering far exceeds the momentum transfer. In nonabelian gauge
theories, it is known that propagators for exchanged gauge bosons
become dressed by a power-like growth in the centre of mass energy, a
phenomenon known as {\it Reggeisation} (see
e.g.~\cite{DelDuca:1995hf}), leading to compact all-order forms for
amplitudes. More recently, the Regge limit has been studied using
Wilson
lines~\cite{Korchemskaya:1994qp,Korchemskaya:1996je,Balitsky:1998kc,Melville:2013qca,Caron-Huot:2013fea},
known factorisation properties of soft and collinear
gluons~\cite{Bret:2011xm,DelDuca:2011ae,DelDuca:2013ara,DelDuca:2014cya},
and effective field theory~\cite{Rothstein:2016bsq}. Reggeisation has
also been examined in (super)-gravity (see
e.g.~\cite{Melville:2013qca} and references therein), where it is
found to be kinematically subleading with respect to other
contributions at high energy.\\

There are a number of motivations for studying the Regge limit in
different theories. In QCD, the physics of Reggeisation (including
non-linear corrections) has potential applications in parton physics
(see
e.g.~\cite{Altarelli:2008aj,White:2006yh,Ciafaloni:2007gf,Bonvini:2016wki}),
multijet
processes~\cite{Andersen:2008ue,Andersen:2008gc,Andersen:2009nu,Andersen:2009he,Andersen:2011hs,Andersen:2011zd,Andersen:2012gk,Andersen:2016vkp,Deak:2009xt,Caporale:2015vya,Vera:2007dr,Vera:2007kn,Bartels:2006hg},
and heavy ion physics~\cite{Mueller:2001fv}. In gravity, the Regge
limit can be used to probe scattering at transplanckian
energies~\cite{'tHooft:1987rb,
  Verlinde:1991iu,Amati:1987uf,Amati:1990xe,Amati:1992zb,Amati:1993tb,Giddings:2010pp},
allowing one to address crucial conceptual issues of quantum gravity,
such as the impact of non-renormalisability, the existence of a
well-defined S-matrix, black hole
physics~\cite{Giddings:2009gj,Giddings:2011xs,Amati:2007ak,Ciafaloni:2015vsa,Ciafaloni:2015xsr},
and connections to string
theory~\cite{D'Appollonio:2010ae,D'Appollonio:2013hja,D'Appollonio:2015xma,D'Appollonio:2015gpa}. As
well as studying each type of field theory individually, there has
been much recent interest in relating (non)-abelian gauge and gravity
theories, motivated in part by the conjectured {\it double copy}
underlying their respective scattering
amplitudes~\cite{Bern:2008qj,Bern:2010ue,Bern:2010yg}. The Regge limit
(as well as more general soft limits) can be used to provide all-order
insights into this
correspondence~\cite{Saotome:2012vy,Vera:2012ds,Oxburgh:2012zr,Melville:2013qca,Johansson:2013nsa},
as well as showing how qualitatively different physics in the two
types of theory are related. To this end, it is useful to develop
languages and techniques for gauge theories and gravity, that make
their common traits particularly clear. \\

An elegant picture for describing the Regge limit of $2\rightarrow 2$
scattering has been provided in
refs.~\cite{Korchemskaya:1994qp,Korchemskaya:1996je}. When the
momentum transfer is much less than the centre of mass energy, the
incoming particles barely glance off each other, and thus follow
approximately straight-line (classical) trajectories. They can thus be
described by Wilson line operators, which take into account the
gauge-covariant phase suffered by each particle as it exchanges soft
(low-momentum) gauge bosons with the
other. References~\cite{Korchemskaya:1994qp,Korchemskaya:1996je}
considered 4-point scattering in QCD, and showed that known properties
of the Regge limit (namely the one-loop Regge trajectory, and infrared
singular part of the two-loop trajectory) can indeed be obtained from
vacuum expectation values of Wilson line operators separated by a
transverse distance $|\vec{z}|$, representing the impact parameter. In
ref.~\cite{Melville:2013qca} this setup was generalised to gravity,
using appropriate gravitational Wilson line operators, introduced and
studied in refs~\cite{Naculich:2011ry,White:2011yy,Miller:2012an} (see
also ref.~\cite{Brandhuber:2008tf}). Existing results regarding the
Regge limits of QCD and gravity were rederived in such a way as to
make the relationship between them especially clear, and the same
method also provided a proof of graviton Reggeisation in $2\rightarrow
n$ processes. \\

The aim of this paper is to extend the results of
ref.~\cite{Melville:2013qca} by systematically including all
corrections that are suppressed by a single power of momentum
transfer. Given the soft nature of the exchanged gauge bosons in the
leading Regge limit (equivalently, the eikonal approximation for the
incoming and outgoing particles), such corrections are referred to as
{\it next-to-soft}, or {\it next-to-eikonal}. There are a number of
motivations for doing this. Firstly, there has recently been a large
amount of attention to amplitudes dressed by additional real emissions
up to next-to-soft level (see
e.g.~\cite{Cachazo:2013hca,Cachazo:2013iea,Strominger:2013jfa,He:2014laa,
  Cachazo:2014fwa,Casali:2014xpa,Schwab:2014xua,Bern:2014oka,He:2014bga,
  Larkoski:2014hta,Cachazo:2014dia,Afkhami-Jeddi:2014fia,Adamo:2014yya,
  Bianchi:2014gla,Bern:2014vva,Broedel:2014fsa,He:2014cra,Zlotnikov:2014sva,
  Kalousios:2014uva,Du:2014eca,Luo:2014wea,White:2014qia}), as well as
previous work from a more phenomenological point of
view~\cite{Low:1958sn,Burnett:1967km,DelDuca:1990gz,Laenen:2008gt,Laenen:2010uz,Bonocore:2015esa,Bonocore:2014wua}. The
present analysis provides an interesting testing ground for these
methods and results. Secondly, corrections to the eikonal
approximation in transplanckian scattering may have a role to play in
furthering our knowledge of quantum gravity (e.g. regarding issues of
black hole production~\cite{Giddings:2010pp,Amati:1992zb}). Thirdly,
by calculating such corrections both in QCD and gravity, one may
further probe the relationship between these two theories.\\

Next-to-soft corrections to the Regge limit in gravity have been
previously considered in detail for massless
particles~\cite{Amati:1987uf,Amati:1990xe,Amati:1992zb,Amati:1993tb},
and also for the case of one particle asymptotically massive, and the
other massless~\cite{Akhoury:2013yua,Bjerrum-Bohr:2016hpa}. Here we
will consider a general situation in which both incoming particles
have (possibly different) masses. The advantage of the massive
situation relative to the completely massless case is that corrections
to the eikonal approximation are enhanced, in that they are suppressed
by fewer powers of the momentum transfer. The kinematic limits adopted
in the previous literature will emerge as special cases. \\

The structure of our paper is as follows. In the following section, we
review the analysis of ref.~\cite{Melville:2013qca} for obtaining the
Regge limit from Wilson lines in position space. In
section~\ref{sec:NE}, we summarise the structure of next-to-soft
corrections, before calculating these in both QCD and gravity. In
section~\ref{sec:discuss} we discuss and interpret our results, before
concluding in section~\ref{sec:conclude}.

\section{Eikonal analysis}
\label{sec:eikonal}

\begin{figure}
\begin{center}
\scalebox{0.8}{\includegraphics{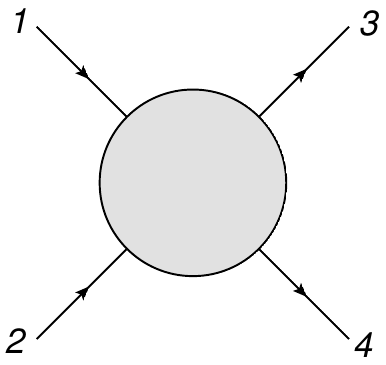}}
\caption{Particle labels used throughout for $2\rightarrow2$ scattering.}
\label{fig:2to2}
\end{center}
\end{figure}
\begin{figure}
\begin{center}
\scalebox{0.8}{\includegraphics{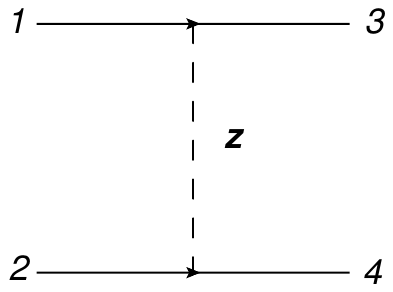}}
\caption{The Regge limit as two Wilson lines separated by a transverse distance $\vec{z}$.}
\label{fig:Wilson}
\end{center}
\end{figure}
Throughout, we consider $2\rightarrow 2$ scattering with momenta
defined as in figure~\ref{fig:2to2}, where we take $m_3=m_1$,
$m_4=m_2$. One may then define the Mandelstam invariants
\begin{equation}
s=(p_1+p_2)^2;\qquad t=(p_1-p_3)^2;\qquad u=(p_1-p_4)^2,
\label{mandies}
\end{equation}
satisfying the momentum conservation constraint
\begin{equation}
s+t+u=2(m_1^2+m_2^2).
\label{momcon}
\end{equation}
When nonzero masses are present, there is a choice regarding how to
define the Regge limit. Following
refs.~\cite{Korchemskaya:1994qp,Korchemskaya:1996je}, we consider the
ordering
\begin{equation}
s\gg m_i^2\gg -t.
\label{Regge}
\end{equation}
When the centre of mass energy dominates the momentum transfer,
particles (1,2) and (3,4) become spacelike collinear to a first
approximation. As discussed in the introduction and in detail in
refs.~\cite{Korchemskaya:1994qp,Korchemskaya:1996je,Melville:2013qca},
one may then represent the incoming and outgoing particles as two
Wilson line operators separated by a transverse vector $\vec{z}$,
where the latter constitutes the impact factor. This setup is depicted
in figure~\ref{fig:Wilson}, and results in the QCD amplitude
\begin{equation}
{\cal A}={\cal A}_E\,{\cal A}_{\rm LO},
\label{Afac}
\end{equation}
where ${\cal A}_{\rm LO}$ is the leading order (Born) amplitude taken
in the Regge limit, which becomes dressed by the eikonal amplitude (in
position space)
\begin{equation}
{\cal A}_E=\left\langle 0\left|\Phi(p_1,0)\Phi(p_2,z)\right|0\right\rangle.
\label{AEdef}
\end{equation}
Here 
\begin{equation}
\Phi(p,x)={\cal P}\exp\left[-ig_s {\bf T}^a p^\mu \int ds A^a_\mu(sp+x)
\right]
\label{phidef}
\end{equation}
is a Wilson line operator describing the emission of soft gluons from
a straightline contour of momentum $p^\mu$, and a constant offset
$x^\mu$. Equation~(\ref{AEdef}) is then a vacuum expectation value of
two Wilson lines, the second of which is displaced with respect to the
first by the constant 4-vector $z$, which is taken to have non-zero
components only in the transverse direction to the incoming
particles. That is, one has~\footnote{We use the metric (+,--,--,--)
  throughout.}
\begin{equation}
z^2=-\vec{z}^2.
\label{z2}
\end{equation}
Were the impact parameter to be zero, eq.~(\ref{AEdef}) would
correspond to the Regge limit of the {\it soft function} describing IR
singularities in a scattering amplitude. As is well known, this soft
function is exactly zero in dimensional regularisation, due to the
cancellation of UV and IR singularities (see e.g.~\cite{Gardi:2009zv}
for a review). The nonzero impact parameter acts as a UV regulator, so
that any remaining singularities are manifestly of infrared origin.\\

\begin{figure}
\begin{center}
\scalebox{0.7}{\includegraphics{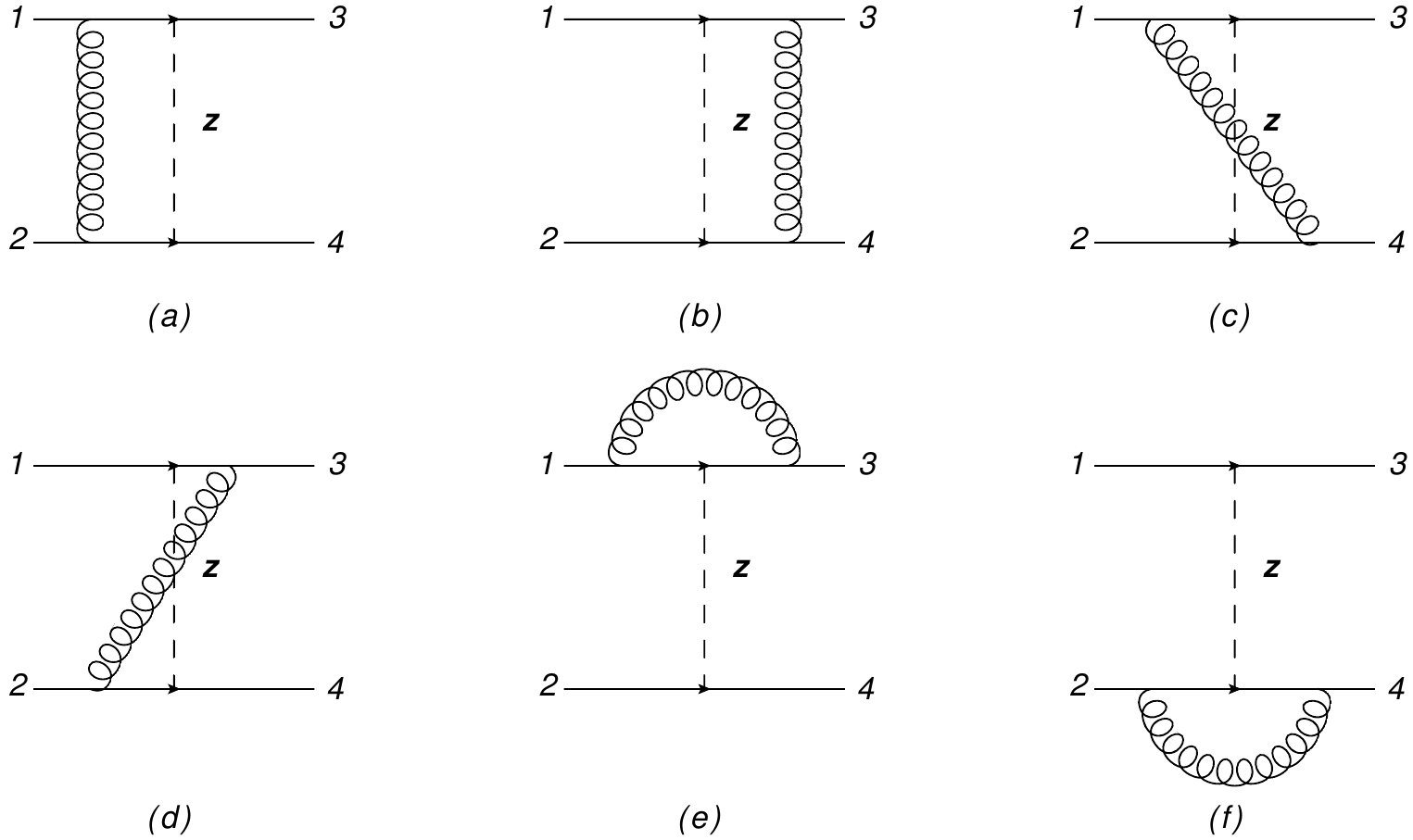}}
\caption{One-loop diagrams entering the calculation of 
eikonal amplitude  ${\cal A}_E$.}
\label{fig:diags}
\end{center}
\end{figure}

One-loop diagrams~\footnote{As in reference~\cite{Melville:2013qca},
  we do not include external self-energies, which lead to constant
  pieces irrelevant for the following discussion.} for the eikonal
amplitude ${\cal A}_E$ are shown in figure~\ref{fig:diags}.
Diagrams (a)--(d) are regulated by the impact parameter, whereas
diagrams (e)--(f) are rendered zero by the presence of an unregulated
UV pole, which cancels the IR behaviour. One may impose a cutoff to
regulate the UV region which, up to logs of the momentum scale choice,
can be chosen to coincide with the same distance scale $|\vec{z}|$
that regulates the remaining graphs. Upon making this choice, the
graphs of figure~\ref{fig:diags} evaluate (in $d=4-2\epsilon$
dimensions, and taking the leading behaviour in $s$)
to~\cite{Melville:2013qca}~\footnote{Reference~\cite{Melville:2013qca}
  treats the case of $m_1=m_2\equiv m$ only. Here we modify the result
  slightly to encompass the unequal mass case.}
\allowdisplaybreaks[0]
\begin{align}
{\cal A}^{(1)}_E=\frac{g_s^2\,\Gamma(1-\epsilon)}{4\pi^{2-\epsilon}}
\frac{(\mu^2\vec{z}^2)^\epsilon}{2\epsilon}&\left\{i\pi\left[{\bf T}_1\cdot
{\bf T}_2+{\bf T}_3\cdot {\bf T}_4\right]
\phantom{\left(\frac{-t}{m_2^2}\right)}\right.\notag\\
&\left.\quad+\log\left(\frac{s}{m_1m_2}\right)
\left[-{\bf T}_1\cdot{\bf T}_2-{\bf T}_3\cdot{\bf T}_4+{\bf T}_1\cdot {\bf T}_4
+{\bf T}_2\cdot {\bf T}_3\right]\right.\notag\\
&\left.\quad+{\bf T}_1\cdot {\bf T}_3
\log\left(-\frac{t}{m_1^2}\right)+{\bf T}_2\cdot {\bf T}_4\log\left(
-\frac{t}{m_2^2}\right)\right\},
\label{AEres}
\end{align}
\allowdisplaybreaks
where ${\bf T}_i$ denotes a colour generator on line $i$, following
the notation of refs.~\cite{Catani:1996jh,Catani:1996vz}, and
satisfying the colour conservation condition
\begin{equation}
{\bf T}_1+{\bf T}_2={\bf T}_3+{\bf T}_4.
\label{colcon}
\end{equation}
Here the $\log(-t/m_i^2)$ terms in eq.~(\ref{AEres}) originate from
diagrams (e)--(f) in figure~\ref{fig:diags}: had we chosen not to
regulate the UV poles in these diagrams, the amplitude would contain
logarithms of $s/(m_1 m_2)$, rather than the expected combination
$s/(-t)$ in the limit of eq.~(\ref{Regge}) (see
e.g. ref~\cite{DelDuca:1995hf}). That this combination indeed results
upon keeping the diagrams involving only a single particle leg can be
seen by defining the quadratic colour operators
\begin{align}
{\bf T}_s^2&=({\bf T}_1+{\bf T}_2)^2=({\bf T}_3+{\bf T}_4)^2,\quad\notag\\
{\bf T}_t^2&=({\bf T}_1-{\bf T}_3)^2=({\bf T}_2-{\bf T}_4)^2,\quad\notag\\
{\bf T}_u^2&=({\bf T}_1-{\bf T}_4)^2=({\bf T}_2-{\bf T}_3)^2,
\label{TSdef}
\end{align}
which, from eq.~(\ref{colcon}), satisfy
\begin{equation}
{\bf T}_s^2+{\bf T}_t^2+{\bf T}_u^2=2C_1+2C_2,\qquad\qquad 
{\bf T}_1^2={\bf T}_3^2 = C_1,\qquad\qquad
{\bf T}_2^2={\bf T}_4^2=C_2.
\label{colcon2}
\end{equation}
Equation~(\ref{AEres}) then becomes
\begin{align}
{\cal A}^{(1)}_E=\frac{g_s^2\,\Gamma(1-\epsilon)}{4\pi^{2-\epsilon}}
\frac{(\mu^2\vec{z}^2)^\epsilon}{2\epsilon}&\left\{i\pi{\bf T}_s^2
+{\bf T}_t^2\log\left(\frac{s}{-t}\right)-i\pi(C_1+C_2)+C_1\log\left(
\frac{-t}{m_1^2}\right)+C_2\log\left(\frac{-t}{m_2^2}\right)
\right\},
\label{AEres2}
\end{align}
thus one indeed sees that the colour non-diagonal terms involve a
logarithm of $s/(-t)$. Given that vacuum expectation values of Wilson
line operators exponentiate (see e.g.~\cite{Gardi:2009zv} for a
review), one may immediately replace eq.~(\ref{AEres2}) with
\begin{align}
{\cal A}_E&=\exp\left\{
\frac{g_s^2\,\Gamma(1-\epsilon)}{4\pi^{2-\epsilon}}
\frac{(\mu^2\vec{z}^2)^\epsilon}{2\epsilon}\left[i\pi{\bf T}_s^2
+{\bf T}_t^2\log\left(\frac{s}{-t}\right)-i\pi(C_1+C_2)
\right.\right.
\notag\\
&\left.\left.\quad\quad\quad\quad\quad\quad
+C_1\log\left(
\frac{-t}{m_1^2}\right)+C_2\log\left(\frac{-t}{m_2^2}\right)\right]
\right\}.
\label{AEres3}
\end{align}
As discussed in
refs.~\cite{Bret:2011xm,DelDuca:2011ae,Melville:2013qca}, the term in
${\bf T}_t^2$ acts as a {\it Reggeisation operator} on the Born
amplitude in eq.~(\ref{Afac}), dressing the exchanged $t$-channel
gluon by a power-like growth in $s/(-t)$, where the associated power
involves the quadratic Casimir of the exchanged particle. The first
term in the exponent in eq.~(\ref{AEres3}) is a pure phase, and is
associated with the formation of bound states in the
$s$-channel~\cite{Brezin:1970zr,Kabat:1992tb}. However, it dominates
only if the quadratic Casimir associated with the $t$-channel exchange
is zero (e.g. for photon exchange), given that the Reggeisation term
is logarithmically enhanced in $s$. Note that eq.~(\ref{AEres3}) has
(logarithmic) singularities as either of the particle masses tends to
zero. These are collinear singularities associated with the incoming
and outgoing particles, and are usually absorbed into {\it impact
  factors} coupling the Reggeised gluon to the upper and lower
particle lines (see e.g.~\cite{DelDuca:1995hf}). \\

It is straightforward to generalise the above analysis to
gravity~\cite{Melville:2013qca}. By analogy with eq.~(\ref{Afac}), one
defines a gravity amplitude
\begin{equation}
{\cal M}={\cal M}_E\,{\cal M}_{\rm LO}.
\label{Mfac}
\end{equation}
Now
\begin{equation}
{\cal M}_E=\left\langle 0\left|\Phi_{g}(p_1,0)
\Phi_{g}(p_2,z)\right|0\right\rangle
\label{MEdef}
\end{equation}
is a vacuum expectation value of two gravitational Wilson line
operators, defined
by~\cite{Naculich:2011ry,White:2011yy,Miller:2012an}
\begin{equation}
\Phi_{g}(p,x)
=\exp\left[\frac{i\kappa}{2} p^\mu\,p^\nu
 \int ds h_{\mu\nu}(sp+x)
\right],\qquad\qquad \kappa^2=32\pi G_N,
\label{phigravdef}
\end{equation}
where $G_N$ is Newton's constant, and we have defined the graviton
according to
\begin{equation}
g_{\mu\nu}=\eta_{\mu\nu}+\kappa h_{\mu\nu}.
\label{gravdef}
\end{equation}
Upon calculating the diagrams of figure~\ref{fig:diags} (including UV
regularisation of graphs (e)--(f) as before), the gravitational
eikonal function in the limit of eq.~(\ref{Regge}) is
\begin{equation}
{\cal M}_E=\exp\left\{-\left(\frac{\kappa}{2}\right)^2
\frac{\Gamma(1-\epsilon)}{4\pi^{2-\epsilon}}\frac{(\mu^2\vec{z}^2)^\epsilon}
{2\epsilon}\left[i\pi s+t\log\left(\frac{s}{-t}\right)\right]
\right\}+{\cal O}(\epsilon^0).
\label{MEres}
\end{equation}
This can also be obtained directly from eq.~(\ref{AEres3}) by making
the replacements
\begin{equation}
g_s\rightarrow\frac{\kappa}{2},\qquad\qquad {\bf T}_s^2\rightarrow s,\qquad\qquad
{\bf T}_t^2\rightarrow t,\qquad\qquad C_i\rightarrow m_i^2,
\label{BCJreplace}
\end{equation}
where terms $\propto m_i^2$ then vanish in the Regge limit~\footnote{
  Reference~\cite{Melville:2013qca} considered the limit $s\gg -t\gg
  m_i^2$ rather than that of eq.~(\ref{Regge}). In either case, one
  may neglect $m_i^2$ relative to $s$.}.  As noted in
ref.~\cite{Melville:2013qca}, these replacements are consistent with
the double copy of
refs.~\cite{Bern:2008qj,Bern:2010ue,Bern:2010yg}. Note that in the
gravity result one may take either mass smoothly to zero, consistent
with the absence of collinear singularities in this
theory~\cite{Weinberg:1965nx,Akhoury:2011kq,Beneke:2012xa}. Due to the
replacements of quadratic colour Casimirs (in QCD) with Mandelstam
invariants (gravity), the $s$-channel phase dominates over the
Reggeisation term, which is power-suppressed. Indeed, the first term
in the exponent of eq.~(\ref{MEres}) is the well-known gravitational
eikonal phase, discussed in detail in refs.~\cite{'tHooft:1987rb,
  Verlinde:1991iu,Amati:1987uf,Amati:1990xe,Amati:1992zb,Amati:1993tb,Giddings:2010pp},
so that in the limit of eq.~(\ref{Regge}) one may write
\begin{equation}
{\cal M}_E=e^{i\chi_{\rm E}},\qquad\qquad \chi_{\rm E}=-\frac{s G_N}{\epsilon}
(\mu^2\vec{z}^2)^\epsilon+{\cal O}(\epsilon^0),
\label{chidef}
\end{equation}
in agreement with e.g. ref.~\cite{Akhoury:2013yua}~\footnote{A similar
  result is provided in ref.~\cite{Kabat:1992tb}, but using a
  fictitious mass for the graviton as an infrared regulator.}.\\

One may connect eqs.~(\ref{Mfac}) and (\ref{chidef}) more directly with the
literature as follows. 
The gravitational Born amplitude consists of a single $t$-channel
graviton exchange, 
which in momentum space gives 
\begin{align}
\tilde{\cal M}_{\rm LO}&=-\frac{i\kappa^2\mu^{2\epsilon}}
{2}\frac{(p_1\cdot p_2)(p_3\cdot p_4)
+(p_1\cdot p_4)(p_2\cdot p_3)-(p_1\cdot p_3)(p_2\cdot p_4)
+m_1^2p_2\cdot p_4+m_2^2p_1\cdot p_3-2m_1^2 m_2^2}{(p_1-p_3)^2}\notag\\\
&=-8\pi i G_N\mu^{2\epsilon}\frac{s^2}{t}+\ldots,
\label{MLO}
\end{align}
where the ellipsis denotes subleading terms as $s\gg -t, m_i^2$. In the
Regge limit, the momentum transfer has components only in the
transverse directions (see e.g. ref.~\cite{Kabat:1992tb}):
\begin{equation}
t\simeq -\vec{q}^2,
\label{tapprox}
\end{equation}
where $\vec{q}$ is the $(d-2)$-dimensional transverse momentum vector
conjugate to the impact parameter $\vec{z}$. The Born amplitude in
impact parameter space is then
\begin{align}
{\cal M}_{\rm LO}&=\int\frac{d^{d-2}\vec{q}}{(2\pi)^{d-2}}
\tilde{\cal M}_{\rm LO}e^{i\vec{q}\cdot\vec{z}}
= 2 i s \chi_{\rm E}
\label{MFourier}
\end{align}
where 
\begin{equation}
\chi_{\rm E}=
-4\pi sG_N\mu^{2\epsilon}
\int\frac{d^{d-2} \vec{k}}{(2\pi)^{d-2}}
\frac{e^{i\vec{k}\cdot \vec{z}}}{-\vec{k}^2} \,.
\label{chiint}
\end{equation}
Carrying out the integral with $d=4 - 2\epsilon$ shows that 
eq.~(\ref{chiint}) is in agreement with  eq.~(\ref{chidef}). 
One may then expand  ${\cal M}_E = e^{i \chi_E}$
and use eq.~(\ref{chiint}) to write~\footnote{Care must be
  taken with combinatorial factors here: in the second line of
  eq.~(\ref{Mexpand}), $n$ represents the number of gluons being
  exchanged, including the Born gluon. An additional factor of
  $n^{-1}$ is then needed in each term due to the fact that the
  symmetric product of integrals introduces an overcounting, by the
  number of ways one can choose which gluon is the Born one.}
\begin{align}
{\cal M}={\cal M}_E\,{\cal M}_{\rm LO}
&=\left[\sum_{m=0}^\infty \frac{(-4\pi i s G_N\mu^{2\epsilon})^m}
{m!}\prod_{i=1}^m\int\frac{d^2\vec{k}_i}{(2\pi)^{d-2}}
\frac{e^{i\vec{k}_i\cdot\vec{z}}}{-\vec{k}_i^2}\right]{\cal M}_{\rm LO}
\notag\\
&=2s\sum_{n=1}^\infty \frac{(-4\pi isG_N\mu^{2\epsilon})^n}{n!}
\prod_{i=1}^n\int
\frac{d^2\vec{k}_i}{(2\pi)^{d-2}}\frac{e^{i\vec{k}_i\cdot\vec{z}}}
{-\vec{k}_i^2}\notag\\
&=2s(e^{i\chi_{\rm E}}-1),
\label{Mexpand}
\end{align}
in agreement with ref.~\cite{Kabat:1992tb}. \\

\section{Beyond the eikonal approximation}
\label{sec:NE}

Having reviewed the eikonal calculation of QCD and gravity scattering
in the Regge limit, we now turn to corrections beyond the leading soft
approximation. To the best of our knowledge, this has not been
previously studied in QCD. In gravity,
refs.~\cite{Amati:1987uf,Amati:1990xe,Amati:1992zb,Amati:1993tb}
considered corrections to the eikonal approximation when both incoming
particles are strictly massless. A dimensional argument can then be
used to show that such corrections are doubly subleading in the impact
factor $|\vec{z}|$. First, one notes that $G_N E$ is the only
classical length scale that one can form, where $E\sim \sqrt{s}$ is
the energy of one of the incoming particles in the centre-of-mass
frame. Then, analyticity of the amplitude requires only integer powers
of $s$, so that the first subleading corrections
\begin{equation}
\sim\frac{G_N^2 s}{|\vec{z}|^2},
\end{equation}
with subsequent corrections also involving only even powers of the
impact parameter. The corrections considered by the above references
thus begin at two-loop order, and are beyond the scope of this
paper.\\

Reference~\cite{Akhoury:2013yua} considered the case of one strictly
massless particle, and the other infinitely massive. In this case one
evades the above dimensional argument due to the presence of an extra
mass scale, such that the first subleading corrections to the eikonal
are ${\cal O}(|\vec{z}|^{-1})$. Here, we will consider the general
situation of two scalar particles with potentially different nonzero
masses, such that the results of~\cite{Akhoury:2013yua} emerge as a
special case~\footnote{The deflection of massless particles with
  different spins was also considered recently in
  ref.~\cite{Bjerrum-Bohr:2016hpa}, with the spinless result agreeing
  with ref.~\cite{Akhoury:2013yua}.}. \\

\begin{figure}
\begin{center}
\scalebox{0.6}{\includegraphics{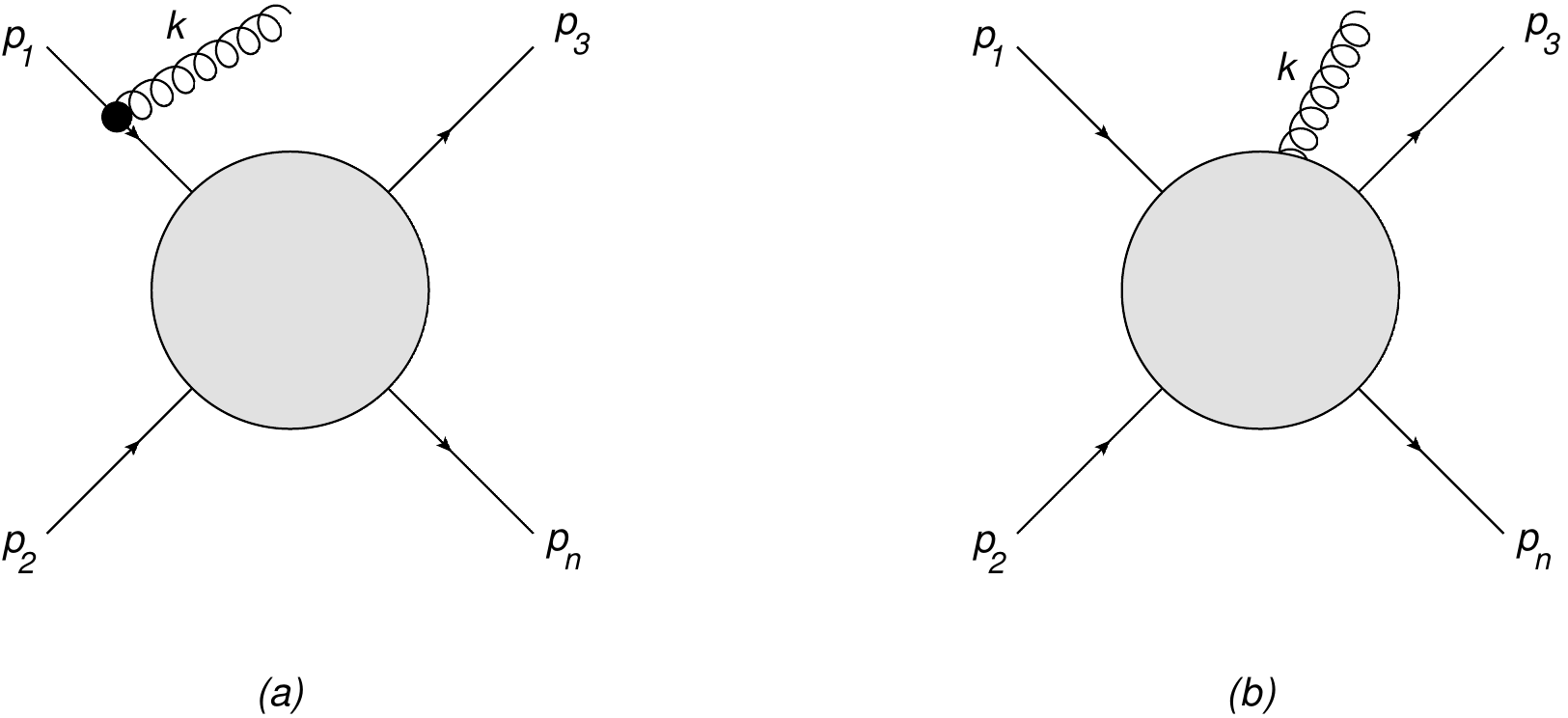}}
\caption{(a) External emisson of a (next-to) soft gluon; (b) Internal
  emission of a soft gluon.}
\label{fig:NEamp}
\end{center}
\end{figure}

To classify next-to-soft corrections, we will use the framework of
refs.~\cite{Laenen:2008gt,White:2011yy} (see also
ref.~\cite{Gellas:1998sh} for similar work in the eikonal
approximation). The starting point is to consider an amplitude with
$n$ external hard particles (i.e. here the four-point amplitude of
figure~\ref{fig:2to2}), to which an additional gluon or graviton
emission is added. There are two possibilities, as shown in
figure~\ref{fig:NEamp}: (i) {\it external emission} contributions, in
which the additional 
boson is emitted from one of the external
legs, and (ii) {\it internal emission contributions}, where the boson
lands inside the nonradiative amplitude. We now deal with each of
these in turn.

\subsection{External emissions in QCD}
\label{sec:extQCD}

As shown in detail in refs.~\cite{Laenen:2008gt,White:2011yy},
external emission contributions are described by generalised Wilson
line operators associated with the hard particle lines. For outgoing
boson momentum $k$, they are given in position space in QCD and
gravity by~\footnote{Note that ref.~\cite{White:2011yy} uses an
  alternative field definition for the graviton. Here we stick to the
  canonical choice of eq.~(\ref{gravdef}).}
\begin{equation}
\Phi_{\rm NE}(p_i,z)={\cal P}
\exp\left\{-ig_s{\bf T}_i\int_0^\infty ds\left[p_{i\mu} 
A^\mu+\frac{i}{2}\partial_\mu A^\mu+\frac{i}{2}tp_{i\mu}
\partial^2A^\mu\right]+{\cal O}(g_s^2)\right\}
\label{QCDWilsongen}
\end{equation}
and
\begin{align}
\Phi_{g,\rm NE}(p_i,z)&=\exp\left\{\frac{i\kappa}{2}
\int_0^\infty ds\left[p_{i\mu}p_{i\nu} 
h^{\mu\nu}+\frac{i}{2}p_{i(\mu}\partial_{\nu)}\left( h^{\mu\nu}
-\frac{h}{2}\eta^{\mu\nu}\right)
+\frac{i}{2}sp_{i\mu}p_{j\nu}\partial^2 h^{\mu\nu}
\right]+{\cal O}(\kappa^2)\right\},
\label{GravWilsongen}
\end{align}
where we have introduced the commonly used notation
\begin{equation}
a^{(\mu}b^{\nu)}=a^\mu b^\nu+a^\nu b^\mu.
\end{equation}
Here $p_i$ is the momentum of the hard emitting particle, whose
trajectory is given, as before, by $x_i^\mu=tp_i^\mu+z$ in general. We
neglect terms quadratic in the coupling constant here, as we will not
need these in the one-loop calculations required for this paper. The
first terms in the exponents of eqs.~(\ref{QCDWilsongen},
\ref{GravWilsongen}) are the usual eikonal Wilson line exponents of
eqs.~(\ref{phidef}). Subsequent terms involve derivatives with respect
to the momentum of the gluon or graviton field, and are thus indeed
subleading in momentum space. They give rise to {\it next-to-eikonal
  Feynman rules} coupling the bosons to the external particle lines,
and we will see explicit examples of their use in the following.\\

\begin{figure}
\begin{center}
\scalebox{0.8}{\includegraphics{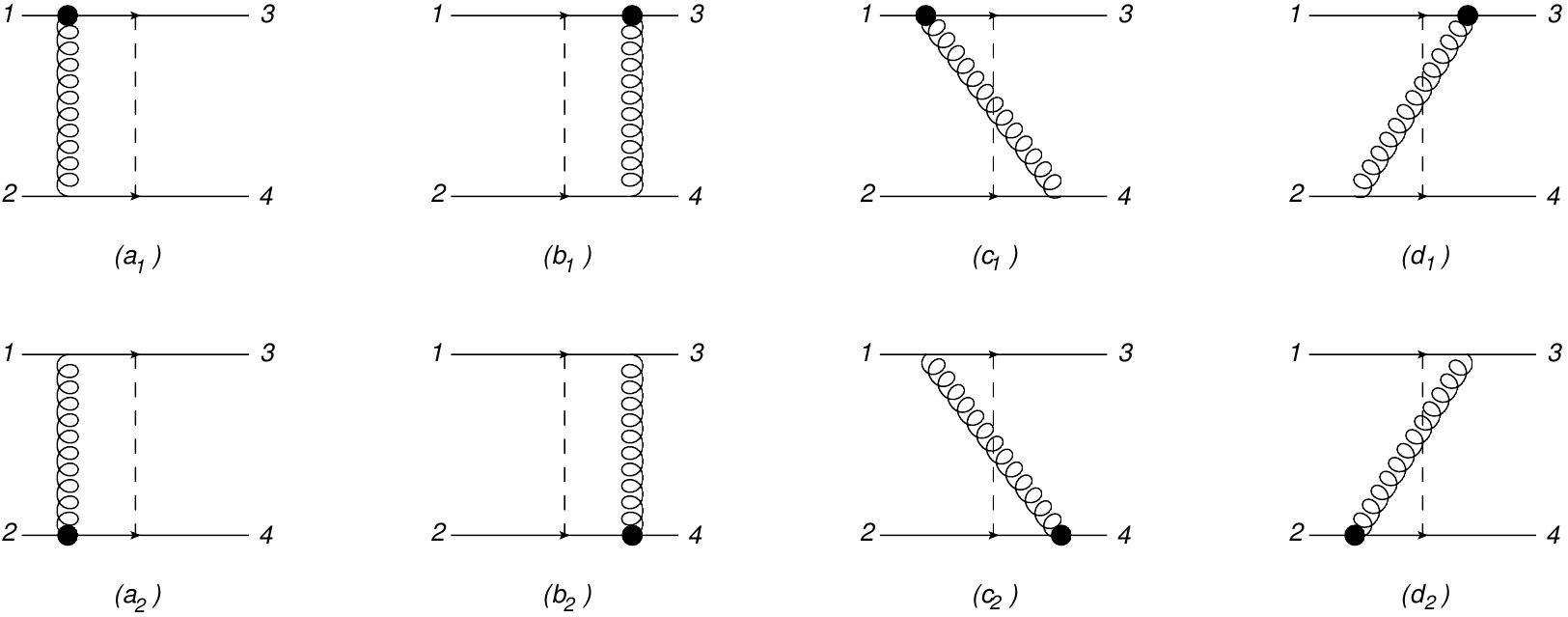}}
\caption{External emission contributions from the generalised Wilson
  line operator of eq.~(\ref{QCDWilsongen}), where $\bullet$
  represents a next-to-soft vertex, and all other vertices are
  eikonal.}
\label{fig:NEdiags}
\end{center}
\end{figure}

\begin{figure}
\begin{center}
\scalebox{0.8}{\includegraphics{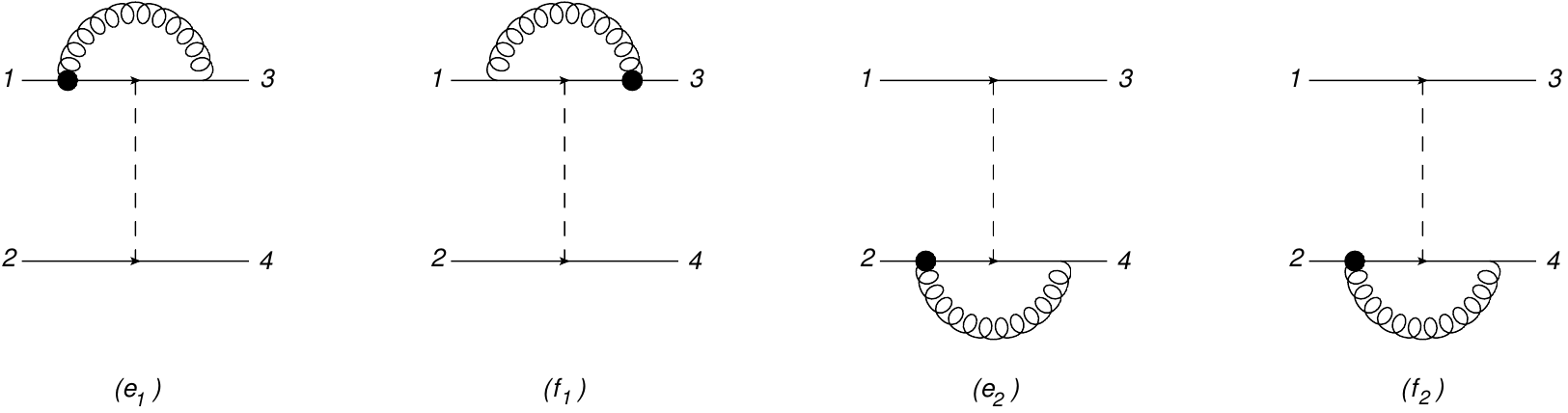}}
\caption{External emission contributions from the generalised Wilson
  line operator of eq.~(\ref{QCDWilsongen}), where $\bullet$
  represents a next-to-soft vertex, and all other vertices are
  eikonal.}
\label{fig:NEdiags2}
\end{center}
\end{figure}

Diagrams contributing at next-to-soft level are shown in
figures~\ref{fig:NEdiags} and~\ref{fig:NEdiags2}.
They can be obtained from the diagrams of figure~\ref{fig:diags} by
replacing at most one eikonal vertex with one of the next-to-soft
Feynman rules from eq.~(\ref{QCDWilsongen}). There are two types,
which in Feynman diagram language have two different origins: the
second term in eq.~(\ref{QCDWilsongen}) arises from corrections to the
numerators associated with gluon emissions on the external lines, and
the third from corrections to the external particle propagator
denominators. In fact, the latter does not contribute, which can be
seen as follows. When embedded in any of the diagrams of
figures~\ref{fig:NEdiags} and~\ref{fig:NEdiags2}, the d'Alembertian
acts on the soft gluon propagator to give
\begin{equation}
\partial^2 D_{\mu\nu}(x-y)=\eta_{\mu\nu}\delta^d(x-y)
\end{equation}
(i.e. the propagator is a Green's function). The right-hand side
implies a non-zero result only if the distance between the two ends of
the soft gluon vanishes. Thus, graphs involving the denominator
correction can potentially contribute only in the absence of a UV
regulator, which acts to remove the short distance region. We will
therefore not have to worry about them in what follows. Note that a
similar conclusion was reached in ref.~\cite{Akhoury:2013yua}, which
separated denominator correction terms into those containing a single
gluon momentum (corresponding to the Feynman rule in
eq.~(\ref{QCDWilsongen})), and those involving a pair of gluon
momenta. The former were argued to vanish for nonzero impact
parameter, as here. The latter are absent in our calculation, as they
correspond to effective Feynman rules involving two or more gauge
bosons, which are absent at one-loop order in the generalised Wilson
line calculation. This corresponds to the fact that such corrections
were also found not to affect the next-to-eikonal phase in
ref.~\cite{Akhoury:2013yua}, due to being higher loop order.\\

It remains to calculate the graphs involving the next-to-soft vertex
in the second term of eq.~(\ref{QCDWilsongen}). As an example, diagram
(b$_1$) is given by
\begin{equation}
{\cal A}_{b_1}=-\frac{ig_s^2(\mu^2)^\epsilon}{2}{\bf T}_3\cdot{\bf T}_4
 \,p_{4\nu}\int_0^\infty ds_3\int_0^\infty ds_4\frac{\partial}
{\partial x_3^\mu}D^{\mu\nu}(x_3-x_4),
\label{diagb1}
\end{equation}
where 
\begin{equation}
x_3^\mu=s_3p_3^\mu+z^\mu,\qquad\qquad x_4^\mu=s_4p_4^\mu,
\end{equation}
and
\begin{equation}
D_{\mu\nu}(x)=-g_{\mu\nu}\frac{\Gamma(d/2-1)}{4\pi^{d/2}}
\left[-x^2+i\varepsilon\right]^{1-d/2}
\label{prop}
\end{equation}
is the position space gluon propagator in $d=4-2\epsilon$ dimensions,
such that
\begin{equation}
\frac{\partial}{\partial x_3^\mu}D^{\mu\nu}(x_3-x_4)=-\frac{\Gamma(d/2)}
{2\pi^{d/2}}(x_3-x_4)^\nu\,[-(x_3-x_4)^2+i\varepsilon]^{-d/2}.
\label{propderiv}
\end{equation}
One may then write eq.~(\ref{diagb1}) as
\begin{equation}
{\cal A}_{b_1}=ig_s^2\mu^{2\epsilon}\frac{\Gamma(d/2)}{4\pi^{d/2}}
{\bf T}_3\cdot {\bf T}_4\, p_{4\mu}\,V_{\rm NE}^\mu(p_3,-p_4),
\label{diagb1b}
\end{equation}
where we have defined the master integral
\begin{equation}
V_{\rm NE}^\mu(\sigma_i p_i,\sigma_j p_j)=\int_0^\infty ds_i\int_0^\infty ds_j
\left(\sigma_i\,s_i\,p_i+\sigma_j\,s_j\,p_j+z\right)^\mu
\left[-(\sigma_ip_i+\sigma_jp_j)^2+\vec{z}^2+i\varepsilon\right]^{-d/2},
\label{VNEdef}
\end{equation}
and $\sigma_{i,j}=\pm1$. One can obtain diagram (b$_1$) by relabelling
$p_3\rightarrow -p_1$, $p_4\rightarrow -p_2$ in
eq.~(\ref{diagb1b}). Similarly, diagram (c$_1$) is given by
\begin{equation}
{\cal A}_{c_1}=ig_s^2\mu^{2\epsilon}\frac{\Gamma(d/2)}{4\pi^{d/2}}
{\bf T}_1\cdot {\bf T}_4\, p_{4\mu}\,V_{\rm NE}^\mu(p_1,p_4),
\label{diagc1}
\end{equation}
with (d$_1$) obtained by relabelling $p_1\rightarrow-p_3$,
$p_4\rightarrow-p_2$. One may also switch momenta to obtain the
diagrams (a$_2$)--(d$_2$), and the integral of eq.~(\ref{VNEdef}) is
calculated in appendix~\ref{app:Vmucalc}. Combining all diagrams, the
total is
\begin{align}
{\cal A}_{a-d}&=\frac{g_s^2\mu^{2\epsilon}}{8\pi^{d/2}}
\Gamma\left(\frac{3}{2}\right)\Gamma\left(\frac{d-3}{2}\right)
|\vec{z}|^{3-d}\left({\bf T}_1\cdot {\bf T}_2+{\bf T}_3\cdot{\bf T}_4
-{\bf T}_1\cdot{\bf T}_4-{\bf T}_2\cdot {\bf T}_3\right)
\left(\frac{1}{m_1}+\frac{1}{m_2}\right)\notag\\
&=-\frac{g_s^2\mu^{2\epsilon}}{8\pi^{d/2}}
\Gamma\left(\frac{3}{2}\right)\Gamma\left(\frac{d-3}{2}\right)
|\vec{z}|^{3-d}\left(\frac{1}{m_1}+\frac{1}{m_2}\right){\bf T}_t^2,
\label{diagsatod}
\end{align}
where we have used the quadratic Casimir operators of
eq.~(\ref{TSdef}).\\

There are a number of noteworthy features of this result. Firstly, it
is IR finite in $d=4$, but contains a pole in $d=3$. The latter is the
analogue of the pole in $d=4$ in the eikonal result of
eq.~(\ref{AEres}). In Feynman diagram language, the (next-to)-eikonal
approximation amounts to linearising denominator factors. At eikonal
level, this introduces a spurious logarithmic UV divergence. Without
any additional regulator, all soft integrals are scaleless, and thus
vanish in dimensional regularisation. The UV pole in eq.~(\ref{AEres})
is, however, regulated by the impact parameter, leaving a remaining IR
pole. At next-to-soft level the story is similar, except for the fact
that going to subleading order in the soft momentum means that the
spurious UV divergence is linear rather than logarithmic. Without an
additional regulator, next-to-soft integrals would be scaleless and
thus vanishing in dimensional regularisation. In this case, however,
one can understand this cancellation as arising between logarithmic
singularities in $d=3$. Regulating the UV divergence with the impact
parameter leaves an (IR) pole in $d=3$, manifest in
eq.~(\ref{diagsatod}).\\

Another property of eq.~(\ref{diagsatod}) is that one cannot take the
massless limit $m_i\rightarrow 0$ for either of the incoming
particles, and the reason for this can again be understood by
comparing with the eikonal result of eq.~(\ref{AEres}). If only
diagrams (a)--(d) in figure~\ref{fig:diags} are included, the one-loop
amplitude contains logarithms of $s/(m_1\, m_2)$, rather than the
conventional combination $s/(-t)$. The remaining diagrams (e) and (f)
are not regulated by the physical impact parameter $\vec{z}$, and
vanish in dimensional regularisation. As discussed in
ref.~\cite{Melville:2013qca} and here in section~\ref{sec:eikonal},
one may choose to also regulate (e) and (f) with the impact parameter,
which amounts to using this as a scale at which to remove the UV
divergence in these diagrams. Whether or not to include diagrams (e)
and (f) thus amounts to a renormalisation scheme choice. The effect of
doing so, as can be seen in eq.~(\ref{AEres3}), is to shift the
logarithms of mass away from the Regge trajectory and into the
colour-diagonal terms. The physical interpretation of these terms is
that they are collinear singularities associated with the incoming and
outgoing particles, where the mass acts as a regulator. The scheme
dependence corresponds to the well-known ambiguity as to whether such
singularities are part of the Regge trajectory, or absorbed into {\it
  impact factors} associated with the upper and lower particle lines
(see e.g. ref.~\cite{DelDuca:1995hf}).\\

The above discussion allows us to interpret the behaviour as
$m_i\rightarrow0$ of eq.~(\ref{diagsatod}): the divergence is
associated with the virtual next-to-soft gluon becoming collinear with
one of the external lines. This divergence is power-like in $d=4$ but
logarithmic in $d=3$, as expected from a divergence which is both
next-to-soft and collinear. Here, as in the eikonal case, we have to
option of including the diagrams (e$_i$) and (f$_i$) in
figure~\ref{fig:NEdiags2}, which amounts to a renormalisation scheme
choice. We instead take the viewpoint of previous
studies~\cite{Amati:1987uf,Amati:1990xe,Amati:1992zb,Amati:1993tb,Kabat:1992tb,Giddings:2010pp,Akhoury:2013yua},
namely that the impact factor implements a physically motivated cutoff
where applicable, and thus only regulate those diagrams in which the
gluons straddle both lines.\\

The power of the generalised Wilson line approach is that, just as in
the eikonal calculation of
refs.~\cite{Korchemskaya:1994qp,Korchemskaya:1996je,Melville:2013qca},
the one-loop amplitude formally
exponentiates~\cite{Laenen:2008gt,White:2011yy}. Keeping only diagrams
(a)--(d) in the eikonal calculation, one may thus write the
generalised Wilson line amplitude as
\begin{align}
{\cal A}_{\rm E +NE}&= \exp\left\{\frac{g^2}{8\pi^{2-\epsilon}}
(\mu^2\vec{z}^2)^\epsilon\left[\frac{\Gamma(1-\epsilon)}{\epsilon}
\left(i\pi ({\bf T}_s^2-C_1-C_2)+{\bf T}_t^2\log\left(\frac{s}{m_1 m_2}
\right)\right)\right.\right.\notag\\
&\left.\left.\qquad\qquad\qquad\qquad\qquad
-\frac{\pi}{2}\frac{{\bf T}_t^2}{|\vec{z}|}
\left(\frac{1}{m_1}+\frac{1}{m_2}\right)+{\cal O}(s^{-1})\right]\right\}
\label{AENEexp}
\end{align}
The colour non-diagonal terms in the eikonal piece (first line)
contain an imaginary piece $\propto {\bf T}_s^2$, and a real part
$\propto {\bf T}_t^2$. As discussed above, the latter corresponds to
the Reggeisation of the gluon, and the former to the eikonal phase
(leading to $s$-channel bound states). In eq.~(\ref{AENEexp}) we see
that at next-to-soft level (second line), there is no imaginary piece,
and thus no next-to-soft correction to the eikonal phase from external
emission contributions. Instead, there is a power-suppressed
correction to the Regge trajectory. This takes the form of pure
collinearly divergent terms, which can be absorbed in the impact
factors associated with the upper and lower lines. \\

Having examined the external emission contributions in QCD, we now
turn to their calculation in gravity.

\subsection{External emissions in gravity}
\label{sec:extgrav}

The diagrams needed for the gravity calculation are again those of
figures~\ref{fig:NEdiags} and~\ref{fig:NEdiags2}, where now we must
use the generalised Wilson line operator of
eq.~(\ref{GravWilsongen}). As in the QCD case, the third term
involving the d'Alembertian operator would contribute only at zero
impact parameter, and thus can be neglected. It is convenient to
rewrite the remaining next-to-soft term via
\begin{equation}
\frac{i\kappa}{2}\int_0^\infty ds\frac{i}{2}p_{i(\mu}\partial_{\nu)}
\left(h^{\mu\nu}-\frac{h}{2}\eta^{\mu\nu}\right)
\rightarrow 
\frac{i\kappa}{2}\int_0^\infty ds\frac{i}{2}p_{i\mu}\partial_{\nu}
\left(\eta^{\mu\alpha}\eta^{\nu\beta}+\eta^{\mu\beta}\eta^{\nu\alpha}
-\eta^{\mu\nu}\eta^{\alpha\beta}\right)h_{\alpha\beta},
\label{NErulegrav}
\end{equation}
where we have used the symmetry of the graviton
$h^{\alpha\beta}=h^{\beta\alpha}$. Diagram (b$_1$) then gives
\begin{align}
{\cal M}_{b_1}&=-\frac{i}{2}\left(\frac{\kappa}{2}\right)^2
\mu^{2\epsilon} p_4^\alpha p_4^\beta p_{3\mu}
\left(\eta^{\mu\sigma}\eta^{\nu\tau}+\eta^{\mu\tau}\eta^{\nu\sigma}
-\eta^{\mu\nu}\eta^{\alpha\beta}\right)
\int_0^\infty ds_3\int_0^\infty ds_4 
\frac{\partial}{\partial x_3^\nu}
\left\langle h_{\sigma\tau}
(x_3)h_{\alpha\beta}(x_4)\right\rangle\notag\\
&=-i\mu^{2\epsilon}
\left(\frac{\kappa}{2}\right)^2\frac{\Gamma(d/2)}{4\pi^{d/2}}
(2p_3\cdot p_4)p_{4\mu}V^\mu_{\rm NE}(p_3,-p_4),
\label{diagb1grav}
\end{align}
where we have used the position-space de Donder gauge graviton propagator
\begin{align}
&\left\langle h_{\sigma\tau}
(x)h_{\alpha\beta}(y)\right\rangle
=P_{\sigma\tau\alpha\beta}
\frac{\Gamma(\frac{d}{2}-1)}{4\pi^{d/2}}
\left[-(x-y)^2+i\varepsilon\right]^{1-d/2},\notag\\
&\quad
P_{\sigma\tau\alpha\beta}=\frac{1}{2}
\left(\eta_{\sigma\alpha}\eta_{\tau\beta}
+\eta_{\sigma\beta}\eta_{\tau\alpha}
-\frac{2}{d-2}\eta_{\sigma\tau}\eta_{\alpha\beta}\right),
\label{gravprop}
\end{align}
as well as the master integral of eq.~(\ref{VNEdef}). The form of
eq.~(\ref{diagb1grav}) is extremely similar to the QCD result of
eq.~(\ref{diagb1b}), and can be obtained from the latter by making the
replacements 
\begin{equation}
g_s\rightarrow\frac{\kappa}{2},\qquad\qquad {\bf T}_i^a\rightarrow p^\mu,
\label{BCJreplace2}
\end{equation}
as well as including an additional factor of 2. As in the eikonal
case, this is precisely consistent with the double
copy~\cite{Bern:2008qj,Bern:2010ue,Bern:2010yg}. The additional factor
is combinatorial in nature, and follows from the fact that numerators
of gravitational integrands result from combining two copies of a
gauge theory numerator. In a given diagram $i$ in which an additional
virtual gluon dresses the Born amplitude (where the latter may be
taken to already be in double copy form), one may expand the extra
contribution to the numerator in the momentum $k$ of the virtual
gluon:
\begin{equation}
n_i=n_i^{(0)}+n_i^{(1)}+{\cal O}(k^2),
\end{equation}
where $n_i^{(m)}$ is the contribution to the numerator at ${\cal
  O}(k^m)$. The gravity numerator for the same graph is then given by
\begin{equation}
n_i n_i=n_i^{(0)}n_i^{(0)}+\left(n_i^{(0)} n_i^{(1)}+n_i^{(1)}n_i^{(0)}
\right)+{\cal O}(k^2),
\end{equation}
and the fact that there are two terms in the ${\cal O}(k)$
contribution is the origin of the additional factor of 2 in
eq.~(\ref{diagb1grav}) relative to the QCD case. One also sees that no
additional factor is present in the leading (eikonal) term, consistent
with the results of ref.~\cite{Melville:2013qca}.\\

The remaining diagrams can be obtained by relabelling
eq.~(\ref{diagb1grav}), or by making the replacements of
eq.~(\ref{BCJreplace2}) and including the above noted factor of 2. The
sum of diagrams (a$_i$)--(d$_i$) is then
\begin{align}
{\cal M}_{a-d}&=\frac{\mu^{2\epsilon}}{4\pi^{d/2}}
\left(\frac{\kappa}{2}\right)^2\Gamma\left(\frac{3}{2}\right)
\Gamma\left(\frac{d-3}{2}\right)|\vec{z}|^{3-d}\left(\frac{1}{m_1}
+\frac{1}{m_2}\right)t.
\label{diagsatodgrav}
\end{align}
Combining this with the eikonal result and exponentiating gives
(c.f. eq.~(\ref{AENEexp}))
\begin{equation}
{\cal M}_{\rm E+NE}=\exp\left\{-\left(\frac{\kappa}{2}\right)^2
\frac{(\mu^2\vec{z}^2)^\epsilon}{8\pi^{2-\epsilon}}
\left[\frac{\Gamma(1-\epsilon)}{\epsilon}\left(i\pi s
+t\log\left(\frac{s}{m_1m_2}\right)\right)
-\frac{\pi t}{|\vec{z}|}\left(\frac{1}{m_1}+\frac{1}{m_2}\right)\right]
\right\}.
\label{MENEexp}
\end{equation}
The effect of the individual colour matrix replacements of
eq.~(\ref{BCJreplace2}) is to replace the $t$-channel quadratic
Casimir appearing in eq.~(\ref{diagsatod}) with the Mandelstam
invariant $t$, as in the previously found eikonal replacements of
eq.~(\ref{BCJreplace}). Similarly to the QCD calculation of the
previous section, one finds a next-to-soft correction to the Regge
trajectory only which, being kinematically subleading in gravity, can
be neglected in the Regge limit. This is consistent with the fact that
external emission contributions (in the present terminology) could be
ignored in ref.~\cite{Akhoury:2013yua}, owing to their being doubly
suppressed in mass and momentum transfer.\\

\subsection{Off-shell internal emissions}
\label{sec:off-shell}

Having calculated the external emission contributions in both QCD and
gravity, we now turn to those soft gluons and gravitons
that arise from inside the
hard interaction. For on-shell bosons, these are given respectively in
QCD and gravity
by~\cite{Low:1958sn,Burnett:1967km,DelDuca:1990gz,Laenen:2008gt,White:2011yy}~\footnote{Our sign in the QCD result matches our convention for the scalar-scalar-gluon vertex (see eq.~(\ref{Vffg}).}
\begin{equation}
{\cal A}^\nu_{\rm int.}=g_s\sum_i {\bf T}_i\left(\eta^{\alpha\nu}
-\frac{\eta_i p_i^\nu k^\alpha}{\eta_i p_i\cdot k+i\varepsilon}\right)
\frac{\partial {\cal A}_n(\{p_i\})}{\partial p_i^\alpha}=
ig_s\sum_i{\bf T}_i^a\frac{L_{\mu\nu}^{(i)}}{p_i\cdot k}
{\cal A}_n(\{p_i\})
\label{internal}
\end{equation}
and
\begin{align}
{\cal M}^{\mu\nu}_{\rm int.}
=-\frac{\kappa}{2}p_i^\mu\sum_i \left(\eta^{\alpha\nu}
-\frac{\eta_i p_i^\nu k^\alpha}{\eta_i p_i\cdot k+i\varepsilon}\right)
\frac{\partial {\cal M}_n(\{p_n\})}{\partial p_i^\alpha}
=-\frac{i\kappa}{2}\sum_i\frac{p_{i\mu} k^\rho L_{\rho\nu}^{(j)}}
{p_j\cdot k}{\cal M}_n(\{p_i\}),
\label{internal2}
\end{align}
where $\eta_i=\pm 1$ according to whether line $i$ is outgoing or
incoming, and we have recognised the orbital angular momentum
generator associated with line $i$:
\begin{equation}
L^{(i)}_{\mu\nu}=x_{i\mu}p_{i\nu}-x_{i\nu}p_{i\mu}
=i\left(p_{i\mu}\frac{\partial}{\partial p_i^\nu}
-p_{i\nu}\frac{\partial}{\partial p_i^\mu}\right).
\label{Ldef}
\end{equation}
This is the same as the total angular
momentum for scalar external particles, and thus eqs.~(\ref{internal},
\ref{internal2}) form a special case of the recently studied {\it
  next-to-soft
  theorems}~\cite{Gross:1968in,Cachazo:2013hca,Cachazo:2013iea,Strominger:2013jfa,He:2014laa,
  Cachazo:2014fwa,Casali:2014xpa,Schwab:2014xua,Bern:2014oka,He:2014bga,
  Larkoski:2014hta,Cachazo:2014dia,Afkhami-Jeddi:2014fia,Adamo:2014yya,
  Bianchi:2014gla,Bern:2014vva,Broedel:2014fsa,He:2014cra,Zlotnikov:2014sva,
  Kalousios:2014uva,Du:2014eca,Luo:2014wea}, as pointed out in more
detail in ref.~\cite{White:2014qia}.  In the present work, all emitted
soft bosons are virtual, and thus off-shell. For the external emission
contributions, this is not a problem, as the generalised Wilson line
operators of eqs.~(\ref{QCDWilsongen}, \ref{GravWilsongen}) are
derived fully generally. Equations~(\ref{internal}, \ref{internal2}),
however, are not guaranteed to work for off-shell bosons. The aim of
this section is to demonstrate that the next-to-soft theorems are
indeed broken by off-shell effects, and to present an alternative way
to calculate the internal emission contributions, motivated by
ref.~\cite{Akhoury:2013yua}.\\

\begin{figure}
\begin{center}
\scalebox{0.5}{\includegraphics{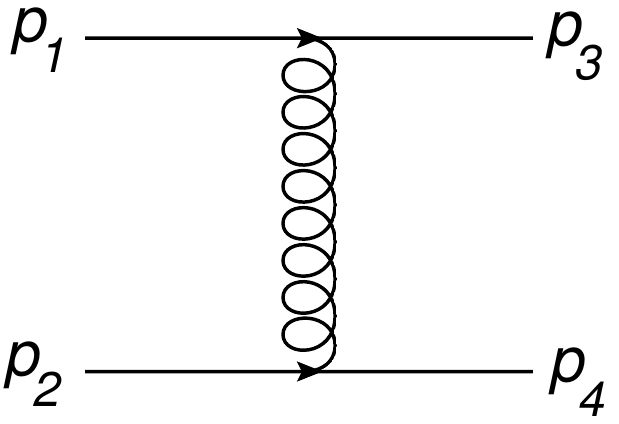}}
\caption{Born diagram for $2\rightarrow2$ scattering in QCD.}
\label{fig:QCDBorn}
\end{center}
\end{figure}

\begin{figure}
\begin{center}
\scalebox{0.7}{\includegraphics{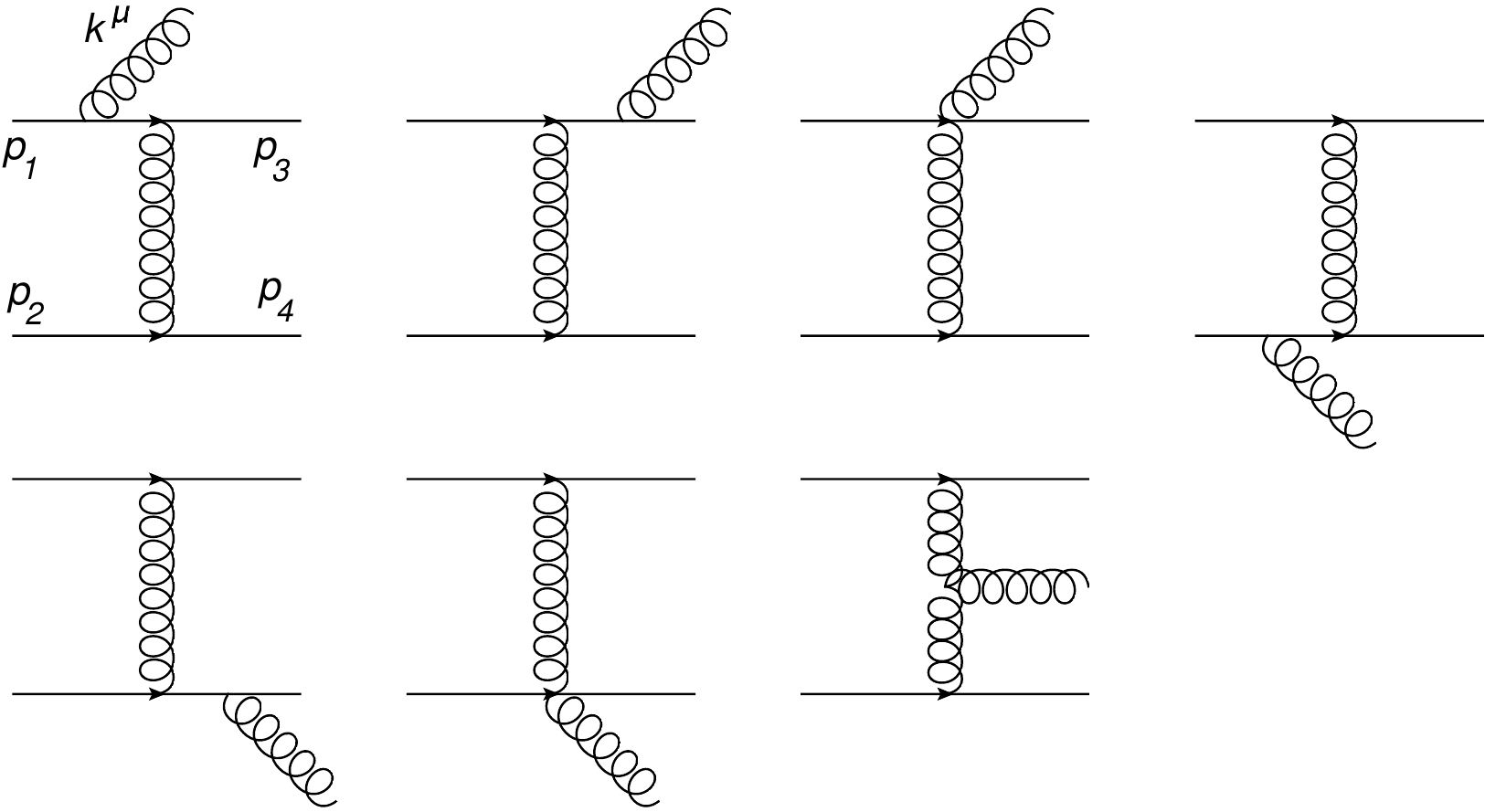}}
\caption{NLO corrections to the Born interaction of 
figure~\ref{fig:QCDBorn}.}
\label{fig:NLOinternal}
\end{center}
\end{figure}

Let us begin by considering the QCD Born interaction for
$2\rightarrow2$ scattering of figure~\ref{fig:QCDBorn}, in which a
hard gluon exchange provides the separation between the incoming
particles that gives rise to the impact factor $\vec{z}$ in the Regge
limit. It is given by
\begin{equation}
\tilde{\cal A}_{\rm LO}=ig_s^2{\bf T}_U^a\,{\bf T}_L^a
 \frac{(p_1+p_3)\cdot (p_2+p_4)}{(p_1-p_3)^2}, 
\label{ALOdef}
\end{equation}
where ${\bf T}^a_{U,L}$ is a colour generator on the upper or
lower line respectively, and the tilde denotes a momentum space
expression.
One may now add an additional off-shell gluon emission which, if not
working in an effective next-to-soft approach, involves the diagrams
of figure~\ref{fig:NLOinternal}. 
These can be evaluated to give~\footnote{We have here suppressed the
  Feynman $i\varepsilon$ prescription for brevity.}
\begin{align}
&\tilde{\cal A}_{\rm NLO}=-ig_s^3\left\{\frac{1}{(p_1-p_3-k)^2}
\left[
{\bf T}_U^a\,{\bf T}_U^b\,{\bf T}_L^a
\frac{(2p_1-k)^\nu}{-2p_1\cdot k+k^2}(p_1+p_3-k)\cdot(p_2+p_4)
\right.\right.\notag\\
&\left.\left.\quad+{\bf T}_U^b\,{\bf T}_U^a\,{\bf T}_L^a
\frac{(2p_3+k)^\nu}{2p_3\cdot k+k^2}(p_1+p_3+k)\cdot(p_2+p_4)
 -\{{\bf T}_U^a,{\bf T}_U^b\}{\bf T}_L^a(p_2+p_4)^\nu\right]\right.\notag\\
&\left.\quad+\frac{1}{(p_1-p_3)^2}\left[
{\bf T}_U^a\,{\bf T}_L^a\,{\bf T}_L^b\frac{(2p_2-k)^\nu}{-2p_2\cdot k+k^2}
(p_1+p_3)\cdot(p_2+p_4-k)\right.\right.\notag\\
&\left.\left.\quad
+{\bf T}_U^a{\bf T}_L^b{\bf T}_L^a\frac{(2p_4+k)^\nu}{2p_4\cdot k+k^2}
(p_1+p_3)\cdot(p_2+p_4+k)
-{\bf T}_U^a\{{\bf T}_L^b,{\bf T}_L^a\}(p_1+p_3)^\nu
\right]\right.\notag\\
&\left.\quad +if^{cba}{\bf T}_U^c\,{\bf T}_L^a\,
\frac{(p_1+p_3)_\mu(p_2+p_4)_\rho}
{(p_1-p_3)^2(p_1-p_3-k)^2}\Big((p_1-p_3+k)^\rho\eta^{\mu\nu}
+(-2k+p_1-p_3)^\mu\eta^{\nu\rho}\right.\notag\\
&\left.\quad+(-2p_1+2p_3+k)^\nu\eta^{\mu\rho}\Big)
\right\}.
\label{ANLO1}
\end{align}
Expanding in the additional gluon momentum $k$ up to next-to-soft
level yields
\begin{align}
\tilde{\cal A}_{\rm NLO}&=-ig_s^3\left\{{\bf T}_U^a\,{\bf T}_U^b\,{\bf T}_L^a
\left[\frac{(p_1+p_3)\cdot(p_2+p_4)}{(p_1-p_3)^2}\left(
-\frac{p_1^\nu}{p_1\cdot k}+\frac{k^\nu}{2p_1\cdot k}-\frac{p_1^\nu k^2}
{2(p_1\cdot k)^2}\right)\right.\right.\notag\\
&\left.\left.+\frac{p_1^\nu}{p_1\cdot k}
\left( \frac{k\cdot(p_2+p_4)}{(p_1-p_3)^2}-\frac{2k\cdot (p_1-p_3)(p_1+p_3)
\cdot(p_2+p_4)}
{(p_1-p_3)^4}\right)\right]\right.\notag\\
&\left.+{\bf T}_U^b\,{\bf T}_U^a\,{\bf T}_L^a\left[\frac{(p_1+p_3)\cdot(p_2+p_4)}
{(p_1-p_3)^2}\left(\frac{p_3^\nu}{p_3\cdot k}+\frac{k^\nu}{2p_3\cdot k}
-\frac{p_3^\nu k^2}{2(p_3\cdot k)^2}\right)\right.\right.\notag\\
&\left.\left.
+\frac{p_3^\nu}{p_3\cdot k}\left(\frac{k\cdot(p_2+p_4)}{(p_1-p_3)^2}
+\frac{2k\cdot (p_1-p_3)(p_1+p_3)\cdot(p_2+p_4)}{(p_1-p_3)^4}\right)
\right]\right.\notag\\
&\left.+{\bf T}_U^a\,{\bf T}_L^a\,{\bf T}_L^b
\left[\frac{(p_1+p_3)\cdot(p_2+p_4)}{(p_1-p_3)^2}
\left(-\frac{p_2^\nu}{p_2\cdot k}
+\frac{k^\nu}{2p_2\cdot k}-\frac{p_2^\nu k^2}{2(p_2\cdot k)^2}\right)
+\frac{p_2^\nu}{p_2\cdot k}\frac{k\cdot(p_1+p_3)}{(p_1-p_3)^2}
\right]\right.\notag\\
&+\left.{\bf T}_U^a\,{\bf T}_L^b\,{\bf T}_L^a\left[
\frac{(p_1+p_3)\cdot(p_2+p_4)}{(p_1-p_3)^2}
\left(\frac{p_4^\nu}{p_4\cdot k}
+\frac{k^\nu}{2p_4\cdot k}-\frac{p_4^\nu k^2}{2(p_4\cdot k)^2}\right)
+\frac{p_4^\nu}{p_4\cdot k}\frac{k\cdot(p_1+p_3)}{(p_1-p_3)^2}
\right]\right.\notag\\
&\left.-\{{\bf T}_U^a,{\bf T}_U^b\}{\bf T}_L^a\frac{(p_2+p_4)^\nu}
{(p_1-p_3)^2}
-{\bf T}_U^a\{{\bf T}_L^a,{\bf T}_L^b\}\frac{(p_1+p_3)^\nu}
{(p_1-p_3)^2}\right.\notag\\
&\left.-2if^{cba}{\bf T}_U^c\,{\bf T}_L^a\frac{(p_1+p_3)\cdot
(p_2+p_4)}{(p_1-p_3)^4}(p_1-p_3)^\nu
\right\}.
\label{ANLO2}
\end{align}
We can recognise some of the terms in this expression (the first group
of terms in each square bracket) as the Born amplitude of
eq.~(\ref{ALOdef}), dressed by eikonal and next-to-eikonal Feynman
rules obtained by Fourier transforming the exponent of
eq.~(\ref{QCDWilsongen}) to momentum space. Thus, these are external
emission contributions, so that the remaining contributions must
correspond to internal emissions. One may then directly check whether
or not they are reproduced from eq.~(\ref{internal}): an 
explicit calculation of the latter gives
\begin{align}
\tilde{\cal A}_{\rm int.}^{b\,\nu}&=-ig_s^3\left\{ {\bf T}_U^a\,{\bf T}_U^b\,
{\bf T}_L^a\,
\frac{p_1^\nu}{p_1\cdot k}\left( \frac{k\cdot(p_2+p_4)}{(p_1-p_3)^2}
-\frac{2k\cdot(p_1-p_3)(p_1+p_3)
\cdot(p_2+p_4)}{(p_1-p_3)^4}\right)\right.\notag\\
&\left.
+{\bf T}_U^b\,{\bf T}_U^a\,{\bf T}_L^a\,
\frac{p_3^\nu}{p_3\cdot k}\left(\frac{k\cdot(p_2+p_4)}{(p_1-p_3)^2}
+\frac{2k\cdot(p_1-p_3)(p_1+p_3)
\cdot(p_2+p_4)}{(p_1-p_3)^4}\right)\right.\notag\\
&\left.+{\bf T}^a_U\,{\bf T}^a_L\,{\bf T}^b_L\,\frac{p_2^\nu}{p_2\cdot k}
\frac{k\cdot(p_1+p_3)}{(p_1-p_3)^2}
+{\bf T}^a_U\,{\bf T}^b_L\,{\bf T}^a_L\,\frac{p_4^\nu}{p_4\cdot k}
\frac{k\cdot(p_1+p_3)}{(p_1-p_3)^2}-{\bf T}_U^a\,\{{\bf T}_L^a,{\bf T}_L^b\}
\frac{(p_1+p_3)^\nu}{(p_1-p_3)^2}\right.\notag\\
&\left.-\{{\bf T}^a_U,{\bf T}^b_U\}{\bf T}^a_L
\frac{(p_2+p_4)^\nu}{(p_1-p_3)^2}
-[{\bf T}_U^b,{\bf T}_U^a]{\bf T}_L^a\,
\frac{2(p_1-p_3)^\nu(p_1+p_3)\cdot(p_2+p_4)}
{(p_1-p_3)^4}
\right\}.
\label{Aintres}
\end{align}
After using the relation
\begin{equation}
[{\bf T}_U^b,{\bf T}_U^a]=if^{bac}{\bf T}_U^c,
\label{Lie}
\end{equation}
eq.~(\ref{Aintres}) precisely reproduces the internal emission terms
in eq.~(\ref{ANLO2}), regardless of the fact that eq.~(\ref{Aintres}) is
manifestly derived for on-shell gluons. It is instructive to classify
the anatomy of this result in more detail. The final three terms in
eq.~(\ref{Aintres}) (those with no explicit dependence on $k$)
originate from the first term in eq.~(\ref{internal}), as must be the
case given that the latter also has no explicit $k$ dependence. In the
full NLO calculation, these correspond to the seagull and three-gluon
vertex graphs, evaluated with $k\rightarrow0$. The remaining terms in
eq.~(\ref{Aintres}) then correspond to the second term in
eq.~(\ref{internal}). Comparison with the full NLO calculation shows
that they have the form of eikonal Feynman rules dressing terms
obtained from the Born interaction by shifting the external momenta in
accordance with the extra gluon emission. This interpretation also
follows directly from the form of eq.~(\ref{internal}), and we will
therefore refer to these contributions as {\it momentum-shift terms}
in what follows.\\

One may carry out a similar analysis for gravity, in which the gluons
in figures~\ref{fig:QCDBorn} and~\ref{fig:NLOinternal} are replaced
with gravitons, and where the Born interaction is now given by
eq.~(\ref{MLO}). The gravitational Feynman rules, including the
three-graviton vertex, may be found in e.g. ref.~\cite{DeWitt:1967uc}
(see also refs.~\cite{Bjerrum-Bohr:2013bxa,BjerrumBohr:2004mz}). Due
to the cumbersome nature of these rules, the full result for the NLO
amplitude, even truncated to next-to-soft order in $k$, is rather
lengthy. We focus only on the non-momentum shift contributions,
stemming from the seagull and three-graviton vertex graphs in
figure~\ref{fig:NLOinternal}. The sum of these contributions as
$k\rightarrow 0$ is given by
\begin{align}
\tilde{\cal  M}^{\mu\nu}&=-\frac{i\kappa^3}{8t}\left[
(m_1^2+m_2^2-s)\left(p_1^{(\mu}p_2^{\nu)}+p_3^{(\mu}p_4^{\nu)}\right)
+(m_1^2+m_2^2-s-t)\left(p_1^{(\mu}p_4^{\nu)}+p_2^{(\mu}p_3^{\nu)}\right)
\right.\notag\\
&\left.
-t (p_1^{(\mu}p_3^{\nu)}+p_2^{(\mu}p_4^{\nu)})
\right]+\kappa\left[\frac{(p_1-p_3)^\mu(p_1-p_3)^\nu}{(p_1-p_3)^2}
+\frac{\eta^{\mu\nu}}{2}\right]\tilde{\cal  M}_{\rm LO}.
\label{total}
\end{align}
As in the QCD case, this should be compared with the first term of
eq.~(\ref{internal2}), and the result is
\begin{align}
\tilde{\cal  M}^{\mu\nu}&=-\frac{i\kappa^3}{8t}\left[
(m_1^2+m_2^2-s)\left(p_1^{(\mu}p_2^{\nu)}+p_3^{(\mu}p_4^{\nu)}\right)
+(m_1^2+m_2^2-s-t)\left(p_1^{(\mu}p_4^{\nu)}+p_2^{(\mu}p_3^{\nu)}\right)
\right.\notag\\
&\left.
-t (p_1^{(\mu}p_3^{\nu)}+p_2^{(\mu}p_4^{\nu)})
\right]+\kappa\left[\frac{(p_1-p_3)^\mu(p_1-p_3)^\nu}{(p_1-p_3)^2}
\right]\tilde{\cal  M}_{\rm LO},
\label{totalderiv}
\end{align}
which agrees with eq.~(\ref{total}) apart from a term involving
$\eta^{\mu\nu}$, and proportional to the Born amplitude. This
contribution vanishes when contracted with a physical graviton
polarisation tensor, and hence eq.~(\ref{internal2}) indeed reproduces
all internal emission contributions provided the additional graviton
emission is on-shell. For off-shell gravitons, however, it constitutes
an explicit breaking of the next-to-soft theorem. The absence of this
breaking in the QCD case is perhaps not surprising - there is no
invariant tensor with one index that could contribute such a term in a
vector theory.\\

\subsection{Seagull and vertex contributions in QCD}
\label{sec:intQCD}

\begin{figure}
\begin{center}
\scalebox{0.8}{\includegraphics{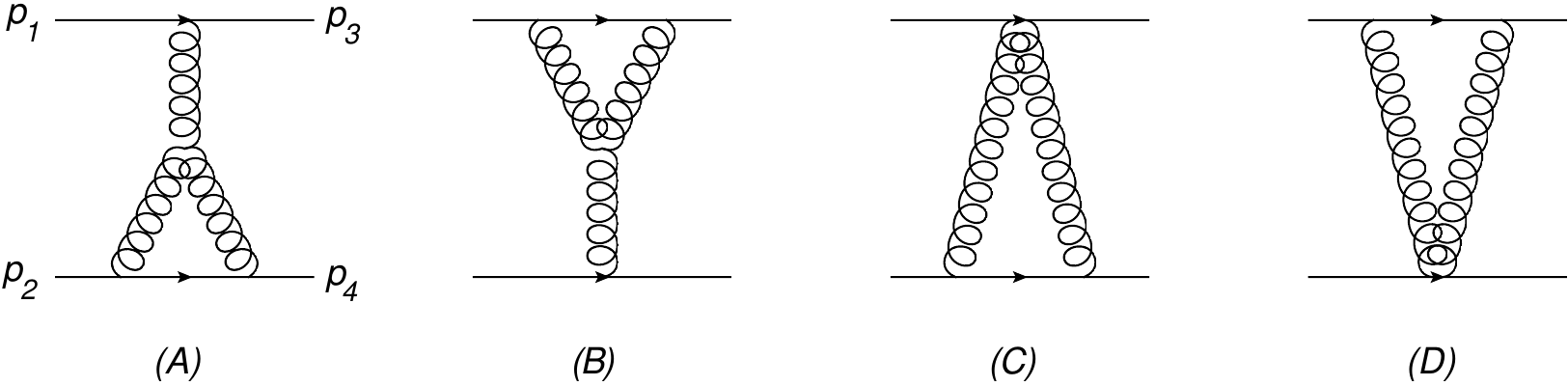}}
\caption{Seagull and triangle diagrams entering the QCD internal
  emission corrections.}
\label{fig:QCDinternal}
\end{center}
\end{figure}

The above analysis implies that we must calculate internal emission
effects by a more direct method. To this end it is useful, as in the
above discussion, to separate the contributions from the seagull and
three-boson vertex graphs, from the momentum-shift contributions
obtained by dressing the shifted Born amplitude with eikonal Feynman
rules. For on-shell emissions, these two types of internal emission
correspond exactly to the first and second terms in
eqs.~(\ref{internal}, \ref{internal2}) respectively, and we begin by
examining the former. The relevant Feynman diagrams are shown in
figure~\ref{fig:QCDinternal}, and we may write the first of these as
\begin{align}
\tilde{\cal A}_{(A)}&=\int\frac{d^d k_1}{(2\pi)^d}
\frac{c_{A}\,n_{A}(\{p_i\},k_1)}{(k_1^2+i\varepsilon)
[(p_1-p_3)^2+i\varepsilon][(p_2+k_1)^2-m_2^2+i\varepsilon]
[(p_1-p_3-k_1)^2+i\varepsilon]},
\label{diagA}
\end{align}
where the colour factor and kinematic numerator are
\begin{align}
c_{A}=f^{abc}\,{\bf T}_U^a\,{\bf T}_L^b\,{\bf T}_L^c
\label{CAQCD}
\end{align}
and
\begin{align}
 n_{A}(\{p_i\},k_1)&=iV_{\phi\phi g}^{\alpha_1}(p_1,-p_3)\,P_{\alpha_1\alpha_2}\,
 V_{ggg}^{\alpha_2\beta_2\gamma_2}[p_1-p_3,-(p_1-p_3-k_1),-k_1]\notag\\
 &\times P_{\gamma_2\gamma_1}\,P_{\beta_2\beta_1}\,V_{\phi\phi g}^{\gamma_1}
 [p_2,-(p_2+k_1)]\,V_{\phi\phi g}^{\beta_1}(p_2+k_1,-p_4)
\label{numAQCD}
\end{align}
respectively. Here 
\begin{equation}
V^\mu_{\phi\phi g}(p_1,p_2)=ig_s(p_1^\mu-p_2^\mu)
\label{Vffg}
\end{equation}
and
\begin{equation}
V^{\alpha\beta\gamma}_{ggg}(p_1,p_2,p_3)=
g_s\left[\eta^{\alpha\beta}(p_1-p_2)^\gamma
+\eta^{\beta\gamma}(p_2-p_3)^\alpha+\eta^{\alpha\gamma}(p_3-p_1)^\beta\right]
\label{Vggg}
\end{equation}
are the scalar-scalar-gluon and three-gluon vertices with all momenta
incoming, and we have defined
\begin{equation}
P_{\alpha\beta}=-i\eta_{\alpha\beta}
\end{equation}
to be the numerator of the Feynman gauge gluon propagator. To extract
the next-to-soft contribution from eq.~(\ref{diagA}), one may
introduce an additional delta function as in
ref.~\cite{Akhoury:2013yua} to rewrite this as
\begin{equation}
\tilde{\cal A}_{(A)}=(2\pi)^d\int\frac{d^d k_1}{(2\pi)^d}
\int\frac{d^d k_2}{(2\pi)^d}\frac{\delta^{(d)}(k_1+k_2-q)\,
c_{A}\,n_{A}}{(k_1^2+i\varepsilon)(k_2^2+i\varepsilon)
[(k_1+k_2)^2+i\varepsilon][(p_2+k_1)^2-m_2^2+i\varepsilon]},
\label{diagA2}
\end{equation}
such that $k_1$ and $k_2$ are now the momenta of the lower two gluons
in figure~\ref{fig:QCDinternal}(A), and we have introduced the
momentum transfer 4-vector (conjugate to $z^\mu$)
\begin{equation}
q^\mu=(p_1-p_3)^\mu.
\label{Deltadef}
\end{equation}
The momenta $k_1$ and $k_2$ are on an equal footing, so that to
isolate next-to-soft contributions, one must expand in both of these
momenta. Returning to the original integral of eq.~(\ref{diagA}), this
can be achieved by writing
\begin{equation}
p_3=p_1-q,\qquad\qquad p_4=p_2+q,
\label{p3p4replace}
\end{equation}
before scaling
\begin{equation}
q\rightarrow \lambda q,\qquad\qquad k_1\rightarrow \lambda k_1,
\label{deltascale}
\end{equation}
and expanding to next-to-soft order in $\lambda$. Finally, one may set
$\lambda\rightarrow 1$. The result may be written
\begin{align}
\tilde{\cal A}_{(A)}=~-~\frac{4ig_s^4 \mu^{4\epsilon}
c_{A}}{q^2}
\Big\{
& q^2(s-m_1^2-m_2^2)S(p_2)
+ \left[ 2 p_2^\mu(s-m_1^2-m_2^2) -4m_2^2 p_1^\mu 
+ 2 m_2^2 q^\mu \right] V_\mu(p_2)
\notag\\
+& [-2p_1^\mu p_2^\nu+(s-m_1^2-m_2^2) \eta^{\mu\nu}]
 T_{\mu\nu}(p_2)  
\Big \},
\label{diagA3}
\end{align}
where we have defined the scalar, vector and tensor integrals
\begin{align}
S(p_i)&=\int\frac{d^d k}{(2\pi)^d}\frac{1}{(k^2+i\varepsilon)
[(q-k)^2+i\varepsilon](2p_i\cdot k+i\varepsilon)};\notag\\
V^\mu(p_i)&=\int\frac{d^d k}{(2\pi)^d}\frac{k^\mu}{(k^2+i\varepsilon)
[(q-k)^2+i\varepsilon](2p_i\cdot k+i\varepsilon)};\notag\\
T^{\mu\nu}(p_i)&=\int\frac{d^d k}{(2\pi)^d}\frac{k^\mu k^\nu}
{(k^2+i\varepsilon)[(q-k)^2+i\varepsilon]
(2p_i\cdot k+i\varepsilon)}.
\label{SVTdef}
\end{align}
We calculate these in appendix~\ref{app:SVTcalc}, and the final result
for diagram (A) is 
\begin{equation}
\tilde{\cal A}_{(A)}=\frac{g_s^4 
c_A(m_1^2+m_2^2-s)}
{16m_2 |\vec{q}|}   
+{\cal O}(\epsilon),
\label{diagAres}
\end{equation}
where $\vec{q}$ is the (two-dimensional) momentum transfer defined in
eq.~(\ref{tapprox}). 
Diagram (B) can be obtained by flipping diagram (A), yielding
\begin{equation}
\tilde{\cal A}_{(B)}= \frac{g_s^4 
c_B(m_1^2+m_2^2-s)}
{16m_1 |\vec{q}|}  
+{\cal O}(\epsilon),
\qquad\qquad  c_B= f^{abc}  {\bf T}_U^b{\bf T}_U^c {\bf T}_L^a.
\label{diagBres}
\end{equation}
Next, one has the seagull graph of figure~\ref{fig:QCDinternal}(C). We
may write this as
\begin{equation}
\tilde{\cal A}_{(C)}=\int\frac{d^d k_1}{(2\pi)^d}\frac{c_C \,n_C(\{p_i\},k_1)}
{(k_1^2+i\varepsilon)[(p_1-p_3-k_1)^2+i\varepsilon]
[(p_2+k_1)^2-m_2^2+i\varepsilon]},
\label{diagC1}
\end{equation}
where the colour factor and kinematic numerator are
\begin{equation}
c_C=\{{\bf T}_U^a,{\bf T}_U^b\}{\bf T}_L^a{\bf T}_L^b
\label{CCQCD}
\end{equation}
and
\begin{equation}
n_C(\{p_i\},k_1)=iV_{\phi\phi g g}^{\alpha_1\beta_1}P_{\alpha_1\alpha_2}
P_{\beta_1\beta_2}V_{\phi\phi g}^{\alpha_2}[p_2,-(p_2+k_1)]
V_{\phi \phi g}^{\beta_2}(p_2+k_1,-p_4),
\label{NCQCD}
\end{equation}
where 
\begin{equation}
V^{\mu\nu}_{\phi\phi g g}=ig_s^2\eta^{\mu\nu}
\label{Vffgg} 
\end{equation}
is the kinematic part of the seagull
vertex. One may expand this according to the procedure of
eqs.~(\ref{p3p4replace}) and~(\ref{deltascale}), and the result is
\begin{align}
\tilde{\cal A}_{(C)}&=-4g_s^4 \,c_C \,m_2^2 \,S(p_2)
=\frac{ig_s^4 \, c_C\,m_2}{8|\vec{q}|}  
+{\cal O}(\epsilon).
\label{diagC2}
\end{align}
Likewise, one has
\begin{align}
\tilde{\cal A}_{(D)}&=\frac{ig_s^4 \,c_D \,m_1}{8|\vec{q}|}
+{\cal O}(\epsilon)
,\qquad\qquad
c_D={\bf T}_U^a\,{\bf T}_U^b \{{\bf T}^a_L,{\bf T}^b_L\}.
\label{diagD}
\end{align}
In order to further interpret these results, it is useful to rewrite
the colour factors in terms of the Born colour factor ${\bf T}_U^a{\bf
  T}_L^a$, and the quadratic Casimir operators of
eq.~(\ref{TSdef}). One has
\begin{align}
{\bf T}_s^2 {\bf T}_U^a{\bf T}_L^a&
=(C_1+C_2){\bf T}_U^a{\bf T}_L^a+2{\bf T}_U^b{\bf T}_U^a{\bf T}_L^b
{\bf T}_L^a;\notag\\
{\bf T}_u^2{\bf T}_U^a{\bf T}_L^a&=(C_1+C_2){\bf T}_U^a{\bf T}_L^a
-2{\bf T}_U^b{\bf T}_U^a{\bf T}_L^a{\bf T}_L^b;\notag\\
{\bf T}_t^2{\bf T}_U^a{\bf T}_L^a&=C_A{\bf T}_U^a{\bf T}_L^a,
\label{TSeq}
\end{align}
such that the various colour factors above can be written
\begin{equation}
c_A=c_B=\frac{i}{2}{\bf T}_t^2{\bf T}_U^a{\bf T}_L^a,\qquad\qquad
c_C=c_D=\frac{({\bf T}_s^2-{\bf T}_u^2)}{2}{\bf T}_U^a{\bf T}_L^a.
\label{colfacs}
\end{equation}
The total contribution from the diagrams of
figure~\ref{fig:QCDinternal} is then
\begin{align}
\tilde{\cal A}_{A-D}&=\frac{ig_s^4}{32|\vec{q}|}\left[
(m_1^2+m_2^2-s)\left(\frac{1}{m_1}+\frac{1}{m_2}\right){\bf T}_t^2
+2(m_1+m_2)({\bf T}_s^2-{\bf T}_u^2)\right]{\bf T}_U^a{\bf T}_L^a
+{\cal O}(\epsilon)
\notag\\
&\rightarrow -\frac{ig_s^4}{|\vec{q}|}  
\frac{s}{32}
\left(\frac{1}{m_1}+\frac{1}{m_2}\right){\bf T}_t^2 {\bf T}_U^a{\bf T}_L^a
+{\cal O}(\epsilon),
\label{MAtoD}
\end{align}
where we have taken the Regge limit in the second line. One may
Fourier transform this result back to impact parameter space, where it
becomes 
\begin{align}
{\cal A}_{A-D}=-\frac{ig_s^4}{|\vec{z}|}\frac{s}{64\pi}
\left(\frac{1}{m_1}+\frac{1}{m_2}\right){\bf T}_t^2 {\bf T}_U^a{\bf T}_L^a 
+{\cal O}(\epsilon).
\label{AAtoDposition}
\end{align}
Comparing this with eq.~(\ref{AENEexp}), we see that the form of
eq.~(\ref{AAtoDposition}) is the same as that of the external emission
correction, namely a $t$-channel Casimir acting on the Born colour
factor, with a real coefficient. Were one able to exponentiate
eq.~(\ref{AAtoDposition}), it would thus correspond to a
power-suppressed correction to the Regge trajectory, rather than the
eikonal phase. For the external emission contributions, exponentiation
follows immediately from the fact that such terms are described by
generalised Wilson line
operators~\cite{Laenen:2008gt,White:2011yy}. For the internal emission
contributions, there is no such argument for exponentiation. However,
one can still choose to exponentiate them: expanding the exponential
will result in higher powers of next-to-soft terms, which are then
higher order in the momentum expansion, and thus of the same formal
accuracy as the non-exponentiated result.

\subsection{Seagull and vertex contributions in gravity}
\label{sec:intgrav}

We may repeat the above analysis for gravity, by replacing the gluons
in figure~\ref{fig:QCDinternal} with gravitons. Given that
intermediate results are a great deal more cumbersome, we here report
the final results only. Diagrams (A) and (C) are found to be given in
momentum space by
\begin{align}
\tilde{\cal M}_{(A)}&=
-\frac{i\kappa^4 m_2}{2048|\vec{q}|}\left[(m_1^2+m_2^2-s)^2
+12m_1^2m_2^2\right]+{\cal O}(\epsilon);\notag\\
\tilde{\cal M}_{(C)}&=\frac{i\kappa^4 m_2}{128|\vec{q}|}\left[m_1^2+m_2^2-s
\right]^2+{\cal O}(\epsilon).
\label{MAMCres}
\end{align}
As before, diagrams (B) and (D) can be obtained by relabelling
$m_1\leftrightarrow m_2$. The sum of all contributions is then
\begin{align}
\tilde{\cal M}_{A-D}&=\frac{i\kappa^4 (m_1+m_2)}{2048|\vec{q}|}
\left[15(m_1^2+m_2^2-s)^2-12m_1^2m_2^2\right]+{\cal O}(\epsilon)\notag\\
&\rightarrow \frac{15i\kappa^4s^2(m_1+m_2)}{2048|\vec{q}|}+{\cal O}(\epsilon),
\label{MAtoDgrav}
\end{align}
where we have taken the Regge limit in the second line. \\

It is interesting to compare eq.~(\ref{MAtoDgrav}) with its
counterpart in QCD, eq.~(\ref{MAtoD}). Up to colour diagonal terms,
the QCD result has a term involving a $t$-channel Casimir that
dominates in the Regge limit, and a suppressed contribution involving
the $s$-channel Casimir (n.b. one may eliminate ${\bf T}_u^2$ in
eq.~(\ref{MAtoD}) using eq.~(\ref{colcon2})). One expects something
like the replacements of eq.~(\ref{BCJreplace}) in moving to the
gravity result, so that the $t$-channel result is subleading, and the
$s$-channel term dominant. Indeed the form of the second term in the
brackets of eq.~(\ref{MAtoD}) is qualitatively the same as
eq.~(\ref{MAtoDgrav}) under eq.~(\ref{BCJreplace}), together with the
additional replacements
\begin{equation}
{\bf T}_{U,L}^a\rightarrow p_{1,2}^\mu,
\label{BCJreplace3}
\end{equation}
consistent with colour generators on the upper and lower lines
corresponding to momenta of these lines in gravity (n.b. one may
equally choose $p_3$ and $p_4$ in this correspondence, given that
$p_1\simeq p_3$ and $p_2\simeq p_4$ up to subleading
corrections). This is analogous to how, at eikonal level, Reggeisation
is the leading effect in QCD, whereas the eikonal phase is more
important in gravity. Note that the coefficient of the $s$-channel
term in QCD is not simply related to that in gravity, which
na\"{i}vely suggests that there is no double copy relationship between
these quantities. This is misleading for a number of reasons. Firstly,
the double copy only formally applies at integrand level, rather than
after integrating over the loop momentum. Secondly, for the double
copy to work for the seagull and vertex contributions, one must choose
a (generalised) gauge such that BCJ duality is manifest in QCD. Here
we have used the Feynman and de Donder gauges in QCD and gravity
respectively, which may obscure a direct double copy. That a double
copy is possible for these graphs, however, follows from the results
of ref.~\cite{Bern:2013yya}.

\subsection{Momentum shift contributions}
\label{sec:momshift}

According to the discussion of section~\ref{sec:off-shell}, the
remaining internal emission contributions comprise the Born
interaction evaluated with shifted momentum, dressed by an additional
eikonal emission. Again regarding as nonzero only those diagrams which
are regulated by the impact factor, the relevant diagrams are those of
figure~\ref{fig:diags}(a)--(d), where the Born amplitude is shifted
appropriately.\\

Focusing first on the case of QCD, the momentum shift contribution
from diagram (a) is given by
\begin{equation}
\tilde{\cal A}_{a}^{\rm mom.}
=\left.-ig_s^2\mu^{2\epsilon}
 {\bf T}_1\cdot{\bf T}_2 p_1\cdot p_2\int\frac{d^d k}{(2\pi)^d}
\frac{\tilde{\cal A}_{\rm LO}(p_1-k,p_2+k)}
{(k^2+i\varepsilon)(-p_1\cdot k+i\varepsilon)
(p_2\cdot k+i\varepsilon)}\right|_{{\cal O}(k)},
\label{momshifta}
\end{equation}
where the numerator contains the Born amplitude of eq.~(\ref{ALOdef}),
and taking the ${\cal O}(k)$ piece isolates the effect of including a
single momentum shift (i.e. terms ${\cal O}(k^2)$ are
next-to-next-to-soft). Substituting eq.~(\ref{ALOdef}) into
eq.~(\ref{momshifta}), the latter becomes
\begin{align}
\tilde{\cal A}_{a}^{\rm mom.}
&=\frac{g_s^4\mu^{4\epsilon}}{2}[{\bf T}_1\cdot{\bf T}_2
{\bf T}_U^a{\bf T}_L^a]
~ p_1\cdot p_2~
(p_1+p_3-p_2-p_4)_\mu
V^\mu_{\rm box}(-p_1,p_2),
\label{momshifta2}
\end{align}
where we have defined the vector box integral
\begin{equation}
V^\mu_{\rm box}=\int\frac{d^d k}{(2\pi)^d}\frac{k^\mu}{(k^2+i\varepsilon)
[(k-q)^2+i\varepsilon](\sigma_i p_i\cdot k+i\varepsilon)
(\sigma_j p_j\cdot k+i\varepsilon)}.
\label{Vboxdef}
\end{equation}
We calculate this integral in appendix~\ref{app:SVTcalc}, and the
result for diagram (a) is (taking the Regge limit)
\begin{equation}
\tilde{\cal A}_{a}^{\rm mom.}=\frac{ig_s^4 s
[{\bf T}_1\cdot {\bf T}_2{\bf T}_U^a{\bf T}_L^a]
(m_1+m_2)}
{16|\vec{q}|m_1m_2}.
\label{momshiftares}
\end{equation}
A similar analysis for diagram (c) yields
\begin{equation}
\tilde{\cal A}_{b}^{\rm mom.}=-\frac{ig_s^4 s
[{\bf T}_1\cdot {\bf T}_4{\bf T}_U^a{\bf T}_L^a]
(m_1+m_2)}
{16|\vec{q}|m_1m_2},
\label{momshiftcres}
\end{equation}
where we have expanded about $d=4$. Diagrams (b) and (d) are equal to
(a) and (c) respectively (n.b. they can be simply obtained by
relabelling masses and colour generators), so that the final result
for the sum of all diagrams is
\begin{align}
\tilde{\cal A}^{\rm mom.}_{a-d}&=-\frac{ig_s^4 s(m_1+m_2)}
{16|\vec{q}|m_1 m_2}{\bf T}_t^2 {\bf T}_U^a{\bf T}_L^a.
\label{momatod}
\end{align}
It is straightforward to carry the above analysis over to gravity
e.g. in eq.~(\ref{momshifta}) one simply replaces the prefactors with
those arising from the gravitational eikonal Feynman rules, and the
Born amplitude in the integrand with that of eq.~(\ref{MLO}). The
final result for the momentum shift contribution upon summing all
diagrams is
\begin{align}
\tilde{\cal M}^{\rm mom.}_{a-d}&=-\frac{i\kappa^4 s^2 t (m_1+m_2)}
{256|\vec{q}|m_1 m_2}.
\label{momatodgrav}
\end{align}
Similarly to the external emisson contributions in
section~\ref{sec:extgrav}, this result can be obtained from the QCD
expression by the replacements of eq.~(\ref{BCJreplace2},
\ref{BCJreplace3}). There is an additional factor of 2 in
eq.~(\ref{momatodgrav}) relative to eq.~(\ref{momatod}) after making
the replacements, which factor has also been explained in
section~\ref{sec:extgrav}.\\

In both QCD and gravity, the momentum shift contributions contain a
$t$-channel Casimir, and thus correspond to shifts in the Regge
trajectory of the gluon / graviton. In gravity, this contribution is
subleading in $t$ and can be discarded. In QCD, the result involves
power-like collinear divergences which can be absorbed into the impact
factors coupling the incoming particles to the Reggeised gluon. That
the momentum shift contributions have the same form as the external
emission contributions of section~\ref{sec:extgrav} is not
surprising. Here we have drawn a distinction between the gluon
entering the Born amplitude, and the external gluons described by
generalised Wilson line operators. Another approach is to consider all
exchanged gluons symmetrically, in which case the momentum shift and
external emission contributions are on an equal footing. The latter
approach is taken in ref.~\cite{Akhoury:2013yua}, which indeed
neglects the momentum shift contributions in gravity as being
subleading.\\

This now completes our calculation of all contributions to
$2\rightarrow 2$ scattering in the high energy limits of QCD and
gravity that are of first subleading order in the momentum
transfer. The more detailed interpretation of these results is the
subject of the following section.

\section{Discussion}
\label{sec:discuss}

In this section, our aim is to draw together the various results of
this paper and discuss their implications in more detail, making
contact with previous calculations in the literature. The complete
next-to-soft corrections in either QCD or gravity are obtained by
summing the external and internal emission contributions. As discussed
above and in refs.~\cite{Laenen:2008gt,White:2011yy}, the former
formally exponentiate, as a direct consequence of being described by
generalised Wilson line operators. The internal emission contributions
can be chosen to exponentiate, given that higher order terms generated
by the exponentiation are progressively subleading in the impact
factor expansion. Upon doing so, all of the QCD contributions in
eqs.~(\ref{diagsatod}, \ref{AAtoDposition}, \ref{momatod}) correspond
to subleading corrections to the Regge trajectory of the gluon. As
already noted in section~\ref{sec:NE}, this correction consists of
purely singular terms as $m_i\rightarrow 0$, associated with the
exchanged gluons becoming collinear with one of the external
lines. These divergences are not problematic in practice, as according
to the Regge limit of eq.~(\ref{Regge}), one cannot take
$m_i\rightarrow 0$ whilst keeping $t$ fixed. One way around this is to
consider the alternative Regge limit
\begin{equation}
s\gg -t\gg m_i^2,
\label{Regge2}
\end{equation}
and to include diagrams such as figure~\ref{fig:diags}(e) and (f),
with a suitable regulator to remove the short-distance singularity. In
the eikonal calculation of ref.~\cite{Melville:2013qca} (reviewed here
in section~\ref{sec:eikonal}), the inclusion of the additional
diagrams explicitly removes collinear singularities from the Regge
trajectory, such that they can be absorbed in so-called {\it impact
  factors} associated with the external lines. Their inclusion in the
Regge trajectory is then a rather unphysical scheme choice, and thus
there is little merit in interpreting the QCD calculation further.\\

The situation in gravity is more interesting. As already remarked in
sections~\ref{sec:extgrav} and~\ref{sec:momshift}, the external
emission and momentum shift contributions are kinematically
subleading, mimicking the suppression of the Regge trajectory at
eikonal level. The only surviving contribution then comes from the
seagull and vertex graphs, and is given in
eq.~(\ref{MAtoDgrav}). Combining this with the eikonal amplitude of
eq.~(\ref{Mexpand}), one may write~\cite{Akhoury:2013yua}
\begin{align}
{\cal M}(\vec{z})&=2s\left[e^{i\chi_{\rm E}(\vec{z})}
\left(1+i\chi_{\rm NE}(\vec{z})\right)-1\right]\notag\\
&=2s\left[e^{i\left(\chi_{\rm E}(\vec{z})
-i\ln[1+i\chi_{\rm NE}(\vec{z})]\right)}
-1\right],
\label{Mexpand2}
\end{align}
where we have defined
\begin{equation}
\chi_{\rm NE}=\frac{15\kappa^4 s^2(m_1+m_2)}{4096\pi|\vec{z}|},
\label{chiNEdef}
\end{equation}
obtained by Fourier transforming eq.~(\ref{MAtoDgrav}) to position
space. In the second line of eq.~(\ref{Mexpand2}) we have written the
NE contribution as the exponential of its own logarithm. Provided that
$\chi_{\rm NE}$ is small, however, one may expand the logarithm so
that the amplitude assumes the simpler form of eq.(\ref{Mexpand}), but
with a total phase
\begin{equation}
\chi=\chi_{\rm E}+\chi_{\rm NE}=G_N s \mu^{2\epsilon}
\left[-\frac{|\vec{z}|^{2\epsilon}}{\epsilon}+\frac{15\pi G_N(m_1+m_2)}
{8|\vec{z}|}\right].
\label{chitotdef}
\end{equation}
This approximation is valid provided the impact parameter is large, or
conversely if the momentum transfer is small relative to the centre of
mass energy. This is precisely the Regge limit of
eq.~(\ref{Regge}). One may now consider the momentum space amplitude
\begin{equation}
\tilde{\cal M}(\vec{q})=\int d^{d-2}\vec{z}\, e^{-i\vec{z}\cdot \vec{q}}
\,{\cal M}(\vec{z}),
\label{Mtildedef}
\end{equation}
where the exponential integral will be dominated by the saddle point,
leading to the stationary phase condition
\begin{equation}
\vec{q}=\frac{\partial \chi}{\partial|\vec{z}|}\frac{\vec{z}}
{|\vec{z}|}.
\label{statphase}
\end{equation}
To interpret this result, let us first consider the case that $m_2\gg
m_1$. This is the situation considered in ref.~\cite{Akhoury:2013yua},
and one may then parametrise
\begin{equation}
p_1^\mu=E_1(1,0,0,1),\qquad\qquad p_2^\mu=(m_2,0,0,0),\qquad\qquad z^\mu=(0,0,|\vec{z}|,0).
\end{equation}
The 4-momentum of the first particle after scattering is
\begin{equation}
{p'_1}^\mu=E_1(1,0,\sin\theta,\cos\theta),
\end{equation}
where $\theta$ is the scattering angle. This in turn implies
\begin{equation}
q^\mu={p'_1}^\mu-p_1^\mu=E_1(0,0,\sin\theta,1-\cos\theta)
\quad\Rightarrow \quad \vec{q}\cdot \vec{z}=-E_1|\vec{z}|\sin\theta
\simeq -E_1|\vec{z}|\theta,
\label{qdef}
\end{equation}
with the small angle approximation justified by the Regge
limit. Equation~(\ref{statphase}) then gives
\begin{align}
\theta &~=~ -\frac{1}{E_1}\frac{\partial \chi}{\partial |\vec{z}|}
~=~\frac{2R_2}{|\vec{z}|}+\frac{15\pi}{16}\left(\frac{R_2}
{|\vec{z}|}\right)^2+\ldots
\label{thetares}
\end{align}
where $R_2=2G_N m_2$ is the Schwarzschild radius associated with the
mass $m_2$, and we have used $s\simeq 2 E_1 m_2$.  The ellipsis
denotes higher order terms in the inverse impact parameter, which mix
with corrections to the next-to-soft approximation and can therefore
be neglected. Equation~(\ref{thetares}) does indeed correspond to the
classical deflection angle experienced by a light test particle
scattering on a black hole (see
e.g. ref.~\cite{D'Appollonio:2010ae}~\footnote{We are very grateful to
  Rodolfo Russo for providing unpublished notes relating to the
  specific case of the Schwarzschild black hole in four dimensions.},
and ref.~\cite{Bjerrum-Bohr:2016hpa} for a recent
derivation). Moreover, the simple form of eq.~(\ref{chiNEdef}) is
independent of whether the mass $m_2$ is small or asymptotically large
relative to $s$. Thus, it applies equally to the case of a test
particle scattering off a black hole, or from a boosted mass, the
extremal case of which is an Aichelburg-Sexl
shockwave~\cite{Aichelburg:1970dh}. This can be further understood
from the fact that at ${\cal O}(G_N)$ one can form two independent
dimensionless combinations from $m_i$, $s$ and $|\vec{z}|$:
\begin{equation}
\frac{G_N m_i}{|\vec{z}|},\qquad\qquad \frac{m_i^2}{s},
\end{equation}
where the first is fixed by the requirement that one expands to
next-to-soft level in the impact parameter only.  In the Regge limit,
the second combination is zero, which uniquely fixes the
next-to-eikonal phase to be linear in the mass of each particle. The
symmetry of eq.~(\ref{chiNEdef}) under interchange of the two masses
shows that the same deflection angle would be experienced by particle
2 treated as a test particle scattering off particle 1. Thus, the
ultimate interpretation of our general next-to-soft calculation is
that it reproduces the two independent classical deflections
experienced by each incoming particle, treated as a test particle in
the field of the other particle.\\

The above discussion relates directly to the investigation of
ref.~\cite{Giddings:2010pp}, which reconsidered transplanckian
scattering in a variety of supersymmetric extensions of gravity,
arguing that additional particle content (and thus the presence or
absence of UV renormalisability) is irrelevant at leading power in the
transplanckian regime. It was pointed out that the complete geometry
corresponding to two colliding shockwaves is not known, and
conjectured that at first subleading level in the momentum expansion
of exchanged gravitons, each incoming shockwave should experience a
classical deflection angle due to the gravitational field of the other
shock. The present analysis precisely confirms this view. It is also
consistent with the known fact that the scattering angle at eikonal
level is the same for a Schwarzschild black hole as for a shockwave
(see e.g.~\cite{Veneziano:2009zz}), and indeed generalises this result
to subleading order in the impact parameter.\\

Some further comments are in order regarding the fact that we have
expanded the logarithm in eq.~(\ref{Mexpand2}). This approximation is
justified when the impact parameter is large, and amounts to
exponentiating the full NE phase. This has been argued to be correct
even for smaller impact parameters, given that at sufficiently large
$s$ the NE correction to the fixed order scattering amplitude violates
unitarity~\cite{D'Appollonio:2010ae}. Reference~\cite{Akhoury:2013yua}
suggested that the seagull and vertex graphs formed part of the
gravitational Wilson line operator, and thus could be
exponentiated. This is not immediately 
borne out in our
approach. However, it may well be that the ${\cal O}(\kappa^2)$ terms
in the generalised Wilson operator of eq.~(\ref{GravWilsongen})
generate multiple copies of the seagull and vertex graphs, in which
case a full exponentiation of these contributions could be formally
proven. 

\section{Conclusion}
\label{sec:conclude}

In this paper, we have examined the high energy (Regge) limit of
$2\rightarrow 2$ scattering in QCD and gravity, extending previous
results to include corrections subleading by a single power of the
impact factor. This generalises previous gravity results for massless
particles~\cite{Amati:1987uf,Amati:1990xe,Amati:1992zb,Amati:1993tb},
and for the case in which only one particle is taken to be
highly massive~\cite{Akhoury:2013yua,Bjerrum-Bohr:2016hpa}. To the best of our
knowledge, no analogous calculations have been carried out in QCD.\\

Our calculational approach builds upon a well-known description of the
Regge limit (at eikonal level) as two Wilson lines separated by a
transverse distance, developed for QCD in
refs.~\cite{Korchemskaya:1994qp,Korchemskaya:1996je}, and applied to
gravity in ref.~\cite{Melville:2013qca}. The generalisation to
next-to-soft level uses the generalised Wilson line approach of
refs.~\cite{Laenen:2008gt,White:2011yy}, which has a number of
significant advantages. Firstly, vacuum expectation values of
generalised Wilson line operators automatically exponentiate,
completely circumventing the combinatorial complexities of
diagrammatic analyses such as that of ref.~\cite{Akhoury:2013yua}
(although, of course, the latter approach remains useful in its own
right). Secondly, the language of generalised Wilson lines reveals
that the calculations in QCD and gravity are extremely similar, even
if the physical interpretation of the results is completely
different. This hints at a deeper underlying relationship between the
two theories, and indeed our results (as discussed in detail
throughout) are entirely consistent with the double copy of
refs.~\cite{Bern:2008qj,Bern:2010ue,Bern:2010yg}.\\

In QCD, we have found a correction to the Regge trajectory of the
gluon, suppressed by a power of the impact parameter, and which is
also purely collinearly singular. This can be removed from the Regge
trajectory by absorbing this correction into impact factors associated
with the incoming particles. However, it would be interesting to see
whether similar methods to those in this paper could be used to study
further power-suppressed terms (in $t/s$) in the Regge limit of
supergravity theories, whose classification remains elusive (see
ref.~\cite{Melville:2013qca} for a recent discussion).\\

In gravity, we have found a general correction to the eikonal phase,
valid for arbitrary masses of the incoming particles. The
interpretation of this correction is that it describes the deflection
angle associated by each particle, considered as a test particle in
the gravitational field of the other. This precisely confirms the
picture conjectured recently in ref.~\cite{Giddings:2010pp}, which
discussed possible interpretations of corrections to eikonal
scattering in supergravity theories.\\

In calculating contributions stemming from soft gluons or gravitons
emanating from inside the hard interaction, we have found that the
gravity next-to-soft theorem of eq.~(\ref{internal2}) is not
sufficient, but must be supplemented by an additional term
proportional to the metric tensor (and which would vanish upon
contraction with a physical polarisation tensor). This seems at odds
with the fact that the result of our gravity calculation is to
reproduce a purely classical effect. It may be that the correction
term is a purely gauge-dependent artifact, but in any case the
generalisation of next-to-soft theorems for off-shell gauge bosons
perhaps deserves further study.\\

Finally, we hope that our paper motivates the further use of
(generalised) gravitational Wilson lines, which have been relatively
unexplored. We believe that they provide an elegant, and panoramic
insight into non-abelian gauge theories and gravity, and our
investigation of further applications is in progress.


\section*{Acknowledgments}

We are very grateful to Pierre Vanhove and Emil Bjerrum-Bohr for
clarification regarding the three-graviton vertex, 
and to Ratin Akhoury and Rodolfo Russo for numerous discussions and
comparisons. 
CDW further thanks Domenico Bonocore, Einan Gardi, Eric
Laenen, Lorenzo Magnea, and Leonardo Vernazza for conversations and
collaboration on related projects. 
AL is supported by a Conacyt studentship, 
and thanks the Centre for Research in String Theory 
at Queen Mary University of London for hospitality. 
SM and CDW are supported by the UK
Science and Technology Facilities Council (STFC). 
SGN is supported by the National Science Foundation
under Grant No.~PHY14-16123.
SGN also gratefully acknowledges sabbatical support
from the Simons Foundation (Grant No.~342554 to Stephen Naculich). 
The authors also thank the 
Michigan Center for Theoretical Physics (University of Michigan)
and the Higgs Centre for Theoretical Physics (University of Edinburgh), 
where part of this research was carried out.

\appendix

\section{Calculation of the master integral $V_{\rm NE}^\mu$}
\label{app:Vmucalc}

In this appendix, we calculate the integral of eq.~(\ref{VNEdef}). One
may first set
\begin{equation}
s_i=\frac{\sqrt{\vec{z}^2}}{m_i}st,\quad 
s_j=\frac{\sqrt{\vec{z}^2}}{m_j}s,
\label{strescale}
\end{equation}
so that eq.~(\ref{VNEdef}) becomes
\begin{align}
V_{\rm NE}^\mu(\sigma_i p_i,\sigma_j p_j)&=
\frac{|\vec{z}|^{3-d}}{m_i m_j}\int_0^\infty dt\int_0^\infty ds
\,s\left(\hat{z}^\mu+st\sigma_i\frac{p_i^\mu}{m_i}
+s\sigma_j\frac{p_j^\mu}{m_j}\right)\notag\\
&\quad\times
\left[1-s^2(t^2+2\sigma t\cosh\gamma_{ij}+1-i\varepsilon)
+i\varepsilon\right]^{-d/2},
\label{VNEcalc1}
\end{align}
where $\hat{z}^\mu=z^\mu/|\vec{z}|$ and $\sigma=\sigma_i\sigma_j$. For
convenience, let us now rewrite this as
\begin{align}
V_{\rm NE}^\mu(\sigma_i p_i,\sigma_j p_j)&=
\frac{|\vec{z}|^{3-d}}{m_i m_j}\left[\hat{z}^\mu V_z+\frac{\sigma_i p_i^\mu}
{m_i}V_i+\frac{\sigma_j p_j^\mu}{m_j}V_j\right].
\label{VNEcalc2}
\end{align}
We will not need to calculate the coefficient $V_z$, due to the fact
that the master integral is only ever contracted with one of the
external lines, and $p_i\cdot z=0$. The coefficient $V_j$ is given by
\begin{align}
V_j&=\int_0^\infty dt\int_0^\infty ds\,s^2\left[
1-s^2(t^2+2\sigma t\cosh\gamma_{ij}
+1-i\varepsilon)+i\varepsilon\right]^{-d/2}\notag\\
&=\sinh\gamma_{ij}\int_{\sigma\coth\gamma_{ij}}^\infty dx
\int_0^\infty ds\,s^2\left[1-s^2\sinh^2\gamma_{ij}(x^2-1-i\varepsilon)
+i\varepsilon\right]^{-d/2},
\label{Vjcalc1}
\end{align}
where we have set $t=x\sinh\gamma_{ij}-\sigma\cosh\gamma_{ij}$ in the
second line. Upon making the substitution
\begin{equation}
s=\sqrt{\frac{u}{1-u}},\quad ds=\frac12\sqrt{\frac{1}{u(1-u)^3}},
\end{equation}
the $s$ integral in eq.~(\ref{Vjcalc1}) becomes 
\begin{align}
&\frac{1}{2}\int_0^1 u^{1/2}(1-u)^{(d-5)/2}\left[1-u(1+\sinh^2\gamma_{ij}
(x^2-1-i\varepsilon)+i\varepsilon)\right]^{-d/2}\notag\\
&=\frac{\Gamma(\frac32)\Gamma(\frac{d}{2}-\frac32)}{2\Gamma(\frac d2)}
{_2}F_1\left(\frac{d}{2},\frac{3}{2};\frac{d}{2};
1+\sinh^2\gamma_{ij}(x^2-1-i\varepsilon)+i\varepsilon\right)\notag\\
&=\frac{\Gamma(\frac32)\Gamma(\frac{d}{2}-\frac32)}{2\Gamma(\frac d2)}
\frac{1}{[-\sinh^2\gamma_{ij}(x^2-1-i\varepsilon)+i\varepsilon]^{3/2}},
\end{align}
where we have used the identity
\begin{equation}
{_2}F_1(a,b;a;z)=(1-z)^{-b}.
\label{hypid}
\end{equation}
Equation~(\ref{Vjcalc1}) now becomes
\begin{align}
V_j&=\frac{\Gamma(\frac32)\Gamma(\frac{d}{2}-\frac32)}
{2\Gamma(\frac{d}{2})\sinh^2\gamma_{ij}}
\int_{\sigma\coth\gamma_{ij}}^\infty\frac{dx}{(1-x^2+i\varepsilon)^{3/2}}.
\label{Vjcalc2}
\end{align}
A careful contour integration gives
\begin{equation}
\int_{\sigma\coth\gamma_{ij}}^\infty \frac{dx}{(1-x^2+i\varepsilon)^{3/2}}
=i(\sigma\cosh\gamma_{ij}-1),
\label{xint}
\end{equation}
so that 
\begin{align}
V_j=\frac{i\Gamma(\frac32)\Gamma(\frac{d}{2}-\frac32)}
{2\Gamma(\frac{d}{2})(1+\sigma\cosh\gamma_{ij})}.
\label{Vjcalc3}
\end{align}
Symmetry of eq.~(\ref{VNEcalc1}) under $i\leftrightarrow j$ implies
that $V_i=V_j$ in eq.~(\ref{VNEcalc2}) (n.b. we have also confirmed
this by explicit calculation). One thus finally obtains
\begin{align}
V_{\rm NE}^\mu(\sigma_i p_i,\sigma_j p_j)&=
\frac{i\Gamma(\frac{d}{2}-\frac32)}{8\pi^{(d-1)/2}}
\frac{|\vec{z}|^{3-d}}{m_i m_j}\left(\frac{\sigma_i p_i^\mu}{m_i}
+\frac{\sigma_j p_j^\mu}{m_j}+\ldots\right)
\frac{1}{(1+\sigma\cosh\gamma_{ij})}.
\label{VNEres}
\end{align}
where the ellipsis denotes terms $\propto z^\mu$.

\section{Calculation of internal emission integrals}
\label{app:SVTcalc}

In this appendix, we calculate the scalar, vector and tensor integrals
of eq.~(\ref{SVTdef}), and the vector box integral of
eq.~(\ref{Vboxdef}). Beginning with the scalar case, one may introduce
Schwinger parameters according to
\begin{equation}
\int_0^\infty ds e^{is(x+i\varepsilon)}=\frac{i}{x+i\varepsilon},
\label{Schwinger}
\end{equation}
yielding~\footnote{Note that we have ignored a term $\sim p_i\cdot q$
  in the exponent of eq.~(\ref{Scalc1}). Keeping this term introduces
  corrections subleading by two powers of $|\vec{q}|$ in the final
  result, which can therefore be neglected.}
\begin{align}
S(p_i)&=i\int\frac{d^d\tilde{k}}{(2\pi)^d}
\int_0^\infty d\alpha_1\int_0^\infty d\alpha_2\int_0^\infty d\alpha_3
\exp\left[i\left((\alpha_1+\alpha_2)\tilde{k}^2-\frac{\alpha_3^2 m_i^2}
{\alpha_1+\alpha_2}+\frac{\alpha_1\alpha_2q^2}{\alpha_1+\alpha_2}\right)
\right],
\label{Scalc1}
\end{align}
where we have also shifted the momentum variable according to
\begin{equation}
\tilde{k}^\mu=k^\mu+\frac{(\alpha_3 p_i-\alpha_2q)^\mu}
{\alpha_1+\alpha_2}.
\label{tildekdef}
\end{equation}
Carrying out the momentum integral gives
\begin{equation}
S(p_i)
=~-~\frac{1}{(4\pi i)^{d/2}}
I\left(\frac{d}{2},0,0\right),
\label{Scalc2}
\end{equation}
where
\begin{equation}
I(l,m,n)=\int_0^\infty d\alpha_1\int_0^\infty d\alpha_2\int_0^\infty
d\alpha_3(\alpha_1+\alpha_2)^{-l}\alpha_2^m\alpha_3^n\exp\left[
-\frac{i}{\alpha_1+\alpha_2}\left(\alpha_3^2m_i^2-\alpha_1\alpha_2q^2\right)
\right]
\label{Idef}
\end{equation}
is a master integral that will be convenient in what follows. The
$\alpha_3$ integral is Gaussian, and can be carried out to give
\begin{equation}
I(l,m,n)=\frac{i^{-(n+1)/2}}{2m_i^{n+1}}\Gamma\left(\frac{1+n}{2}\right)
\int_0^\infty d\alpha_1\int_0^\infty d\alpha_2\,\alpha_2^m
(\alpha_1+\alpha_2)^{-l+(n+1)/2}\exp\left[\frac{i\alpha_1\alpha_2 q^2}
{\alpha_1+\alpha_2}\right].
\label{Icalc1}
\end{equation}
One may now transform 
\begin{equation}
\alpha_1=\alpha x,\quad \alpha_2=\alpha(1-x),\quad d\alpha_1d\alpha_2=
\alpha d\alpha dx,
\label{alphatrans}
\end{equation}
followed by
\begin{equation}
\alpha=\frac{i\beta}{x(1-x)q^2}
\label{alphatrans2}
\end{equation}
to get
\begin{align}
I(l,m,n)&=\frac{i^{-n+l-m-3}}{2m_i^{n+1}}\Gamma\left(\frac{1+n}{2}\right)
(-q^2)^{l-(n+1)/2-m-2}
\int_0^\infty d\beta \beta^{m+1-l+(n+1)/2}e^{-\beta}\notag\\
&\quad\times\int_0^1 dx x^{l-(n+1)/2-m-2}(1-x)^{l-(n+1)/2-2}\notag\\
&=\frac{i^{-n+l-m-3}}{2m_i^{n+1}}\frac{\Gamma(\frac{1+n}{2})
\Gamma(m-l+\frac{n}{2}+\frac{5}{2})\Gamma(l-\frac{n}{2}-\frac{3}{2})
\Gamma(l-\frac{n}{2}-m-\frac{3}{2})}{\Gamma(2l-n-m-3)}
|\vec{q}|^{2l-n-2m-5},
\label{Icalc2}
\end{align}
where we have defined the square of the two-dimensional momentum
transfer via (c.f. eq.~(\ref{tapprox}))
\begin{equation}
q^2\simeq-\vec{q}^2.
\label{vecdeldef}
\end{equation}
Substituting eq.~(\ref{Icalc2}) into eq.~(\ref{Scalc2}), the final
result for the scalar integral is
\begin{equation}
S(p_i)=-\frac{i\sqrt{\pi}}{2(4\pi)^{d/2}}\frac{|\vec{q}|^{d-5}}{m_i}
\frac{\Gamma\left(\frac{5-d}{2}\right)\Gamma^2\left(\frac{d-3}{2}\right)}
{\Gamma(d-3)}.
\label{Sres}
\end{equation}
One may carry out the momentum integrals for the vector and tensor
cases in a similar manner. They are given in terms of the master
integral of eq.~(\ref{Idef}) as follows:
\begin{align}
V^\mu(p_i)&=
~-~\frac{1}{(4\pi i)^{d/2}}
\left[
- p_i^\mu I\left(\frac{d}{2}+1,0,1\right)
+ q^\mu I\left(\frac{d}{2}+1,1,0\right)
\right]
;\notag\\
T^{\mu\nu}(p_i)&=
~-~\frac{1}{(4\pi i)^{d/2}}
\left[p_i^\mu p_i^\nu
I\left(\frac{d}{2}+2,0,2\right)-q^{(\mu}p_i^{\nu)}
I\left(\frac{d}{2}+2,1,1\right)\right.\notag\\
&\left.\quad\qquad\qquad\qquad+q^\mu q^\nu 
I\left(\frac{d}{2}+2,2,0\right)+\frac{i}{2}\eta^{\mu\nu}
I\left(\frac{d}{2}+1,0,0\right)\right].
\end{align}
Let us now turn to the vector box integral of
eq.~(\ref{Vboxdef}). Introducing Schwinger parameters, this is given
by
\begin{align}
V_{\rm box}^\mu&=4\int\frac{d^dk}{(2\pi)^d}
\int_0^\infty d\alpha_1\int_0^\infty d\alpha_2
\int_0^\infty d\alpha_3\int_0^\infty d\alpha_4
\,k^\mu\notag\\
&\quad\times
\exp\left[i\alpha_1 k^2+i\alpha_2(q-k)^2+2i\sigma_i\alpha_3 p_i\cdot k
+2i\sigma_j\alpha_4p_j\cdot k-\sum_i\alpha_i\epsilon\right]\notag\\
&=-\frac{4i^{1-d/2}}{(4\pi)^{d/2}}
\int_0^\infty d\alpha_1\int_0^\infty d\alpha_2
\int_0^\infty d\alpha_3\int_0^\infty d\alpha_4
\left(\sigma_i\alpha_3 p_i+\sigma_j\alpha_4 p_j-\alpha_2q\right)^\mu
(\alpha_1+\alpha_2)^{-1-d/2}\notag\\
&\quad\times
\exp\left[\frac{i\alpha_1\alpha_2 q^2}{\alpha_1+\alpha_2}
+\frac{i(-\alpha_3^2 m_i^2-\alpha_4 m_j^2-2\alpha_3\alpha_4\sigma m_i m_j
\cosh\gamma_{ij}+i\epsilon)}{\alpha_1+\alpha_2}\right],
\label{Vboxcalc1}
\end{align}
where we have carried out the momentum integration in the second
equality, defined $\sigma=\sigma_i\sigma_j$, and absorbed positive
definite factors into $\varepsilon$ where necessary. Here the term in
$q^\mu$ may be ignored, as it will vanish upon contraction with
any external momenta. For the term in $p_i^\mu$, one may rescale
$\alpha_3\rightarrow \alpha_3\sqrt{\alpha_1+\alpha_2}/m_i$,
$\alpha_4\rightarrow \alpha_4\sqrt{\alpha_1+\alpha_2}/m_j$, then make
the transformations of eqs.~(\ref{alphatrans}, \ref{alphatrans2}) to
carry out the ($\alpha_1$, $\alpha_2$) integrals, leaving
\begin{align}
\left.V^\mu_{\rm box}\right|_{p_i^\mu}&=
-\frac{4i^{-3/2}}{(4\pi)^{d/2}}\frac{\sigma_i p_i^\mu}{m_i^2 m_j}
\frac{\Gamma(\frac{5}{2}-\frac{d}{2})
\Gamma^2(\frac{d}{2}-\frac{3}{2})}{\Gamma(d-3)}|\vec{q}|^{d-5}
\int_0^\infty d\alpha_3\int_0^\infty d\alpha_4\, \alpha_3\notag\\
&\quad\times
\exp\left[i(-\alpha_3^2-\alpha_4^2-2\alpha_3\alpha_4\sigma
\cosh\gamma_{ij}+i\varepsilon)\right].
\label{Vboxcalc2}
\end{align}
After setting $\alpha_4\rightarrow\alpha_3\alpha_4$, the $\alpha_3$
integral may be carried out to give
\begin{align}
\left.V^\mu_{\rm box}\right|_{p_i^\mu}&=
-\frac{\sqrt{\pi}}{(4\pi)^{d/2}}\frac{\sigma_i p_i^\mu}{m_i^2 m_j}
\frac{\Gamma(\frac{5}{2}-\frac{d}{2})
\Gamma^2(\frac{d}{2}-\frac{3}{2})}{\Gamma(d-3)}|\vec{q}|^{d-5}
\int_0^\infty \frac{d\alpha_4}
{(-1-\alpha_4^2-2\alpha_4\sigma
\cosh\gamma_{ij}+i\varepsilon)^{3/2}},
\label{Vboxcalc3}
\end{align}
and the transformation
$\alpha_4=x\sinh\gamma_{ij}-\sigma\cosh\gamma_{ij}$ subsequently yields
\begin{align}
\left.V^\mu_{\rm box}\right|_{p_i^\mu}&=
-\frac{\sqrt{\pi}}{(4\pi)^{d/2}}\frac{\sigma_i p_i^\mu}
{m_i^2 m_j\sinh^2\gamma_{ij}}
\frac{\Gamma(\frac{5}{2}-\frac{d}{2})
\Gamma^2(\frac{d}{2}-\frac{3}{2})}{\Gamma(d-3)}|\vec{q}|^{d-5}
\int_{\sigma\coth\gamma_{ij}}^\infty [1-x^2+i\varepsilon]^{-3/2}.
\label{Vboxcalc4}
\end{align}
The $x$ integral has already been carried out in eq.~(\ref{xint}).
Furthermore, symmetry allows the coefficient of $p_j^\mu$ in
eq.~(\ref{Vboxcalc1}) to be straightforwardly obtained from that of
$p_i^\mu$. The final result for the box integral is
\begin{align}
V_{\rm box}^\mu=-\frac{i\sqrt{\pi}}{(4\pi)^{d/2}}\frac{1}{m_i m_j
(1+\sigma\cosh\gamma_{ij})}\frac{\Gamma(\frac{5}{2}-\frac{d}{2})
\Gamma^2(\frac{d}{2}-\frac{3}{2})}{\Gamma(d-3)}|\vec{q}|^{d-5}
\left(\frac{\sigma_ip_i^\mu}{m_i}+\frac{\sigma_j p_j^\mu}{m_j}\right)+\ldots,
\label{Vboxres}
\end{align}
where the ellipsis denotes the term in $q^\mu$ that can be
ignored.

\bibliography{refs.bib}

\providecommand{\href}[2]{#2}\begingroup\raggedright\begin{thebibliography}{100}

\bibitem{DelDuca:1995hf}
V.~Del~Duca, ``{An introduction to the perturbative QCD pomeron and to jet
  physics at large rapidities},''
\href{http://www.arXiv.org/abs/hep-ph/9503226}{{\tt hep-ph/9503226}}.

\bibitem{Korchemskaya:1994qp}
I.~Korchemskaya and G.~Korchemsky, ``{High-energy scattering in QCD and cross
  singularities of Wilson loops},'' {\em Nucl.Phys.} {\bf B437} (1995)
  127--162,
\href{http://www.arXiv.org/abs/hep-ph/9409446}{{\tt hep-ph/9409446}}.

\bibitem{Korchemskaya:1996je}
I.~A. Korchemskaya and G.~P. Korchemsky, ``{Evolution equation for gluon Regge
  trajectory},'' {\em Phys. Lett.} {\bf B387} (1996) 346--354,
\href{http://www.arXiv.org/abs/hep-ph/9607229}{{\tt hep-ph/9607229}}.

\bibitem{Balitsky:1998kc}
I.~Balitsky, ``{Factorization for high-energy scattering},'' {\em Phys. Rev.
  Lett.} {\bf 81} (1998) 2024--2027,
\href{http://www.arXiv.org/abs/hep-ph/9807434}{{\tt hep-ph/9807434}}.

\bibitem{Melville:2013qca}
S.~Melville, S.~G. Naculich, H.~J. Schnitzer, and C.~D. White, ``{Wilson line
  approach to gravity in the high energy limit},'' {\em Phys. Rev.} {\bf D89}
  (2014), no.~2, 025009,
\href{http://www.arXiv.org/abs/1306.6019}{{\tt 1306.6019}}.

\bibitem{Caron-Huot:2013fea}
S.~Caron-Huot, ``{When does the gluon reggeize?},'' {\em JHEP} {\bf 05} (2015)
  093,
\href{http://www.arXiv.org/abs/1309.6521}{{\tt 1309.6521}}.

\bibitem{Bret:2011xm}
V.~Del~Duca, C.~Duhr, E.~Gardi, L.~Magnea, and C.~D. White, ``{An infrared
  approach to Reggeization},'' {\em Phys.Rev.} {\bf D85} (2012) 071104,
\href{http://www.arXiv.org/abs/1108.5947}{{\tt 1108.5947}}.

\bibitem{DelDuca:2011ae}
V.~Del~Duca, C.~Duhr, E.~Gardi, L.~Magnea, and C.~D. White, ``{The Infrared
  structure of gauge theory amplitudes in the high-energy limit},'' {\em JHEP}
  {\bf 1112} (2011) 021,
\href{http://www.arXiv.org/abs/1109.3581}{{\tt 1109.3581}}.

\bibitem{DelDuca:2013ara}
V.~Del~Duca, G.~Falcioni, L.~Magnea, and L.~Vernazza, ``{High-energy QCD
  amplitudes at two loops and beyond},'' {\em Phys. Lett.} {\bf B732} (2014)
  233--240,
\href{http://www.arXiv.org/abs/1311.0304}{{\tt 1311.0304}}.

\bibitem{DelDuca:2014cya}
V.~Del~Duca, G.~Falcioni, L.~Magnea, and L.~Vernazza, ``{Analyzing high-energy
  factorization beyond next-to-leading logarithmic accuracy},'' {\em JHEP} {\bf
  02} (2015) 029,
\href{http://www.arXiv.org/abs/1409.8330}{{\tt 1409.8330}}.

\bibitem{Rothstein:2016bsq}
I.~Z. Rothstein and I.~W. Stewart, ``{An Effective Field Theory for Forward
  Scattering and Factorization Violation},'' {\em JHEP} {\bf 08} (2016) 025,
\href{http://www.arXiv.org/abs/1601.04695}{{\tt 1601.04695}}.

\bibitem{Altarelli:2008aj}
G.~Altarelli, R.~D. Ball, and S.~Forte, ``{Small x Resummation with Quarks:
  Deep-Inelastic Scattering},'' {\em Nucl. Phys.} {\bf B799} (2008) 199--240,
\href{http://www.arXiv.org/abs/0802.0032}{{\tt 0802.0032}}.

\bibitem{White:2006yh}
C.~D. White and R.~S. Thorne, ``{A Global Fit to Scattering Data with NLL BFKL
  Resummations},'' {\em Phys. Rev.} {\bf D75} (2007) 034005,
\href{http://www.arXiv.org/abs/hep-ph/0611204}{{\tt hep-ph/0611204}}.

\bibitem{Ciafaloni:2007gf}
M.~Ciafaloni, D.~Colferai, G.~P. Salam, and A.~M. Stasto, ``{A Matrix
  formulation for small-x singlet evolution},'' {\em JHEP} {\bf 08} (2007) 046,
\href{http://www.arXiv.org/abs/0707.1453}{{\tt 0707.1453}}.

\bibitem{Bonvini:2016wki}
M.~Bonvini, S.~Marzani, and T.~Peraro, ``{Small-$x$ resummation from HELL},''
  {\em Eur. Phys. J.} {\bf C76} (2016) 597,
\href{http://www.arXiv.org/abs/1607.02153}{{\tt 1607.02153}}.

\bibitem{Andersen:2008ue}
J.~R. Andersen and C.~D. White, ``{A New Framework for Multijet Predictions and
  its application to Higgs Boson production at the LHC},'' {\em Phys. Rev.}
  {\bf D78} (2008) 051501,
\href{http://www.arXiv.org/abs/0802.2858}{{\tt 0802.2858}}.

\bibitem{Andersen:2008gc}
J.~R. Andersen, V.~Del~Duca, and C.~D. White, ``{Higgs Boson Production in
  Association with Multiple Hard Jets},'' {\em JHEP} {\bf 02} (2009) 015,
\href{http://www.arXiv.org/abs/0808.3696}{{\tt 0808.3696}}.

\bibitem{Andersen:2009nu}
J.~R. Andersen and J.~M. Smillie, ``{Constructing All-Order Corrections to
  Multi-Jet Rates},'' {\em JHEP} {\bf 01} (2010) 039,
\href{http://www.arXiv.org/abs/0908.2786}{{\tt 0908.2786}}.

\bibitem{Andersen:2009he}
J.~R. Andersen and J.~M. Smillie, ``{The Factorisation of the t-channel Pole in
  Quark-Gluon Scattering},'' {\em Phys. Rev.} {\bf D81} (2010) 114021,
\href{http://www.arXiv.org/abs/0910.5113}{{\tt 0910.5113}}.

\bibitem{Andersen:2011hs}
J.~R. Andersen and J.~M. Smillie, ``{Multiple Jets at the LHC with High Energy
  Jets},'' {\em JHEP} {\bf 06} (2011) 010,
\href{http://www.arXiv.org/abs/1101.5394}{{\tt 1101.5394}}.

\bibitem{Andersen:2011zd}
J.~R. Andersen, L.~Lonnblad, and J.~M. Smillie, ``{A Parton Shower for High
  Energy Jets},'' {\em JHEP} {\bf 07} (2011) 110,
\href{http://www.arXiv.org/abs/1104.1316}{{\tt 1104.1316}}.

\bibitem{Andersen:2012gk}
J.~R. Andersen, T.~Hapola, and J.~M. Smillie, ``{W Plus Multiple Jets at the
  LHC with High Energy Jets},'' {\em JHEP} {\bf 09} (2012) 047,
\href{http://www.arXiv.org/abs/1206.6763}{{\tt 1206.6763}}.

\bibitem{Andersen:2016vkp}
J.~R. Andersen, J.~J. Medley, and J.~M. Smillie, ``{$Z/\gamma^{∗}$ plus
  multiple hard jets in high energy collisions},'' {\em JHEP} {\bf 05} (2016)
  136,
\href{http://www.arXiv.org/abs/1603.05460}{{\tt 1603.05460}}.

\bibitem{Deak:2009xt}
M.~Deak, F.~Hautmann, H.~Jung, and K.~Kutak, ``{Forward Jet Production at the
  Large Hadron Collider},'' {\em JHEP} {\bf 09} (2009) 121,
\href{http://www.arXiv.org/abs/0908.0538}{{\tt 0908.0538}}.

\bibitem{Caporale:2015vya}
F.~Caporale, G.~Chachamis, B.~Murdaca, and A.~S. Vera,
  ``{Balitsky-Fadin-Kuraev-Lipatov Predictions for Inclusive Three Jet
  Production at the LHC},'' {\em Phys. Rev. Lett.} {\bf 116} (2016), no.~1,
  012001,
\href{http://www.arXiv.org/abs/1508.07711}{{\tt 1508.07711}}.

\bibitem{Vera:2007dr}
A.~Sabio~Vera and F.~Schwennsen, ``{Azimuthal decorrelation of forward jets in
  Deep Inelastic Scattering},'' {\em Phys. Rev.} {\bf D77} (2008) 014001,
\href{http://www.arXiv.org/abs/0708.0549}{{\tt 0708.0549}}.

\bibitem{Vera:2007kn}
A.~Sabio~Vera and F.~Schwennsen, ``{The Azimuthal decorrelation of jets widely
  separated in rapidity as a test of the BFKL kernel},'' {\em Nucl. Phys.} {\bf
  B776} (2007) 170--186,
\href{http://www.arXiv.org/abs/hep-ph/0702158}{{\tt hep-ph/0702158}}.

\bibitem{Bartels:2006hg}
J.~Bartels, A.~Sabio~Vera, and F.~Schwennsen, ``{NLO inclusive jet production
  in $k_T$-factorization},'' {\em JHEP} {\bf 11} (2006) 051,
\href{http://www.arXiv.org/abs/hep-ph/0608154}{{\tt hep-ph/0608154}}.

\bibitem{Mueller:2001fv}
A.~H. Mueller, ``{Parton saturation: An Overview},'' in {\em {QCD perspectives
  on hot and dense matter. Proceedings, NATO Advanced Study Institute, Summer
  School, Cargese, France, August 6-18, 2001}}, pp.~45--72.
\newblock 2001.
\newblock
\href{http://www.arXiv.org/abs/hep-ph/0111244}{{\tt hep-ph/0111244}}.
\newblock

\bibitem{'tHooft:1987rb}
G.~'t~Hooft, ``Graviton dominance in ultrahigh-energy scattering,'' {\em
  Phys.Lett.} {\bf B198} (1987)
61--63.

\bibitem{Verlinde:1991iu}
H.~L. Verlinde and E.~P. Verlinde, ``{Scattering at Planckian energies},'' {\em
  Nucl.Phys.} {\bf B371} (1992) 246--268,
\href{http://www.arXiv.org/abs/hep-th/9110017}{{\tt hep-th/9110017}}.

\bibitem{Amati:1987uf}
D.~Amati, M.~Ciafaloni, and G.~Veneziano, ``{Classical and Quantum Gravity
  Effects from Planckian Energy Superstring Collisions},'' {\em
  Int.J.Mod.Phys.} {\bf A3} (1988)
1615--1661.

\bibitem{Amati:1990xe}
D.~Amati, M.~Ciafaloni, and G.~Veneziano, ``Higher order gravitational
  deflection and soft bremsstrahlung in {P}lanckian energy superstring
  collisions,'' {\em Nucl.Phys.} {\bf B347} (1990)
550--580.

\bibitem{Amati:1992zb}
D.~Amati, M.~Ciafaloni, and G.~Veneziano, ``{Planckian scattering beyond the
  semiclassical approximation},'' {\em Phys.Lett.} {\bf B289} (1992)
87--91.

\bibitem{Amati:1993tb}
D.~Amati, M.~Ciafaloni, and G.~Veneziano, ``{Effective action and all order
  gravitational eikonal at Planckian energies},'' {\em Nucl.Phys.} {\bf B403}
  (1993)
707--724.

\bibitem{Giddings:2010pp}
S.~B. Giddings, M.~Schmidt-Sommerfeld, and J.~R. Andersen, ``{High energy
  scattering in gravity and supergravity},'' {\em Phys.Rev.} {\bf D82} (2010)
  104022,
\href{http://www.arXiv.org/abs/1005.5408}{{\tt 1005.5408}}.

\bibitem{Giddings:2009gj}
S.~B. Giddings and R.~A. Porto, ``{The Gravitational S-matrix},'' {\em Phys.
  Rev.} {\bf D81} (2010) 025002,
\href{http://www.arXiv.org/abs/0908.0004}{{\tt 0908.0004}}.

\bibitem{Giddings:2011xs}
S.~B. Giddings, ``{The gravitational S-matrix: Erice lectures},'' {\em Subnucl.
  Ser.} {\bf 48} (2013) 93--147,
\href{http://www.arXiv.org/abs/1105.2036}{{\tt 1105.2036}}.

\bibitem{Amati:2007ak}
D.~Amati, M.~Ciafaloni, and G.~Veneziano, ``{Towards an S-matrix description of
  gravitational collapse},'' {\em JHEP} {\bf 02} (2008) 049,
\href{http://www.arXiv.org/abs/0712.1209}{{\tt 0712.1209}}.

\bibitem{Ciafaloni:2015vsa}
M.~Ciafaloni, D.~Colferai, and G.~Veneziano, ``{Emerging Hawking-Like Radiation
  from Gravitational Bremsstrahlung Beyond the Planck Scale},'' {\em Phys. Rev.
  Lett.} {\bf 115} (2015), no.~17, 171301,
\href{http://www.arXiv.org/abs/1505.06619}{{\tt 1505.06619}}.

\bibitem{Ciafaloni:2015xsr}
M.~Ciafaloni, D.~Colferai, F.~Coradeschi, and G.~Veneziano, ``{Unified limiting
  form of graviton radiation at extreme energies},'' {\em Phys. Rev.} {\bf D93}
  (2016), no.~4, 044052,
\href{http://www.arXiv.org/abs/1512.00281}{{\tt 1512.00281}}.

\bibitem{D'Appollonio:2010ae}
G.~D'Appollonio, P.~Di~Vecchia, R.~Russo, and G.~Veneziano, ``{High-energy
  string-brane scattering: Leading eikonal and beyond},'' {\em JHEP} {\bf 11}
  (2010) 100,
\href{http://www.arXiv.org/abs/1008.4773}{{\tt 1008.4773}}.

\bibitem{D'Appollonio:2013hja}
G.~D'Appollonio, P.~Vecchia, R.~Russo, and G.~Veneziano, ``{Microscopic unitary
  description of tidal excitations in high-energy string-brane collisions},''
  {\em JHEP} {\bf 11} (2013) 126,
\href{http://www.arXiv.org/abs/1310.1254}{{\tt 1310.1254}}.

\bibitem{D'Appollonio:2015xma}
G.~D'Appollonio, P.~Di~Vecchia, R.~Russo, and G.~Veneziano, ``{A microscopic
  description of absorption in high-energy string-brane collisions},'' {\em
  JHEP} {\bf 03} (2016) 030,
\href{http://www.arXiv.org/abs/1510.03837}{{\tt 1510.03837}}.

\bibitem{D'Appollonio:2015gpa}
G.~D'Appollonio, P.~Di~Vecchia, R.~Russo, and G.~Veneziano, ``{Regge behavior
  saves String Theory from causality violations},'' {\em JHEP} {\bf 05} (2015)
  144,
\href{http://www.arXiv.org/abs/1502.01254}{{\tt 1502.01254}}.

\bibitem{Bern:2008qj}
Z.~Bern, J.~Carrasco, and H.~Johansson, ``{New Relations for Gauge-Theory
  Amplitudes},'' {\em Phys.Rev.} {\bf D78} (2008) 085011,
\href{http://www.arXiv.org/abs/0805.3993}{{\tt 0805.3993}}.

\bibitem{Bern:2010ue}
Z.~Bern, J.~J.~M. Carrasco, and H.~Johansson, ``{Perturbative Quantum Gravity
  as a Double Copy of Gauge Theory},'' {\em Phys.Rev.Lett.} {\bf 105} (2010)
  061602, \href{http://www.arXiv.org/abs/1004.0476}{{\tt 1004.0476}}.

\bibitem{Bern:2010yg}
Z.~Bern, T.~Dennen, Y.-t. Huang, and M.~Kiermaier, ``{Gravity as the Square of
  Gauge Theory},'' {\em Phys.Rev.} {\bf D82} (2010) 065003,
  \href{http://www.arXiv.org/abs/1004.0693}{{\tt 1004.0693}}.

\bibitem{Saotome:2012vy}
R.~Saotome and R.~Akhoury, ``{Relationship Between Gravity and Gauge Scattering
  in the High Energy Limit},'' {\em JHEP} {\bf 1301} (2013) 123,
\href{http://www.arXiv.org/abs/1210.8111}{{\tt 1210.8111}}.

\bibitem{Vera:2012ds}
A.~S. Vera, E.~S. Campillo, and M.~A. Vazquez-Mozo, ``{Color-Kinematics Duality
  and the Regge Limit of Inelastic Amplitudes},'' {\em J. High Energy Phys.}
  {\bf 04} (2013) 086,
\href{http://www.arXiv.org/abs/1212.5103}{{\tt 1212.5103}}.

\bibitem{Oxburgh:2012zr}
S.~Oxburgh and C.~White, ``{BCJ duality and the double copy in the soft
  limit},'' {\em JHEP} {\bf 1302} (2013) 127,
\href{http://www.arXiv.org/abs/1210.1110}{{\tt 1210.1110}}.

\bibitem{Johansson:2013nsa}
H.~Johansson, A.~Sabio~Vera, E.~Serna~Campillo, and M.~Ã. Vázquez-Mozo,
  ``{Color-Kinematics Duality in Multi-Regge Kinematics and Dimensional
  Reduction},'' {\em JHEP} {\bf 10} (2013) 215,
\href{http://www.arXiv.org/abs/1307.3106}{{\tt 1307.3106}}.

\bibitem{Naculich:2011ry}
S.~G. Naculich and H.~J. Schnitzer, ``{Eikonal methods applied to gravitational
  scattering amplitudes},'' {\em JHEP} {\bf 1105} (2011) 087,
\href{http://www.arXiv.org/abs/1101.1524}{{\tt 1101.1524}}.

\bibitem{White:2011yy}
C.~D. White, ``{Factorization Properties of Soft Graviton Amplitudes},'' {\em
  JHEP} {\bf 1105} (2011) 060, \href{http://www.arXiv.org/abs/1103.2981}{{\tt
  1103.2981}}.

\bibitem{Miller:2012an}
D.~Miller and C.~White, ``{The Gravitational cusp anomalous dimension from AdS
  space},'' {\em Phys.Rev.} {\bf D85} (2012) 104034,
\href{http://www.arXiv.org/abs/1201.2358}{{\tt 1201.2358}}.

\bibitem{Brandhuber:2008tf}
A.~Brandhuber, P.~Heslop, A.~Nasti, B.~Spence, and G.~Travaglini, ``{Four-point
  Amplitudes in N=8 Supergravity and Wilson Loops},'' {\em Nucl.Phys.} {\bf
  B807} (2009) 290--314,
\href{http://www.arXiv.org/abs/0805.2763}{{\tt 0805.2763}}.

\bibitem{Cachazo:2013hca}
F.~Cachazo, S.~He, and E.~Y. Yuan, ``{Scattering of Massless Particles in
  Arbitrary Dimensions},'' {\em Phys. Rev. Lett.} {\bf 113} (2014), no.~17,
  171601,
\href{http://www.arXiv.org/abs/1307.2199}{{\tt 1307.2199}}.

\bibitem{Cachazo:2013iea}
F.~Cachazo, S.~He, and E.~Y. Yuan, ``{Scattering of Massless Particles:
  Scalars, Gluons and Gravitons},'' {\em JHEP} {\bf 07} (2014) 033,
\href{http://www.arXiv.org/abs/1309.0885}{{\tt 1309.0885}}.

\bibitem{Strominger:2013jfa}
A.~Strominger, ``{On BMS Invariance of Gravitational Scattering},'' {\em JHEP}
  {\bf 07} (2014) 152,
\href{http://www.arXiv.org/abs/1312.2229}{{\tt 1312.2229}}.

\bibitem{He:2014laa}
T.~He, V.~Lysov, P.~Mitra, and A.~Strominger, ``{BMS supertranslations and
  Weinberg's soft graviton theorem},'' {\em JHEP} {\bf 05} (2015) 151,
\href{http://www.arXiv.org/abs/1401.7026}{{\tt 1401.7026}}.

\bibitem{Cachazo:2014fwa}
F.~Cachazo and A.~Strominger, ``{Evidence for a New Soft Graviton Theorem},''
\href{http://www.arXiv.org/abs/1404.4091}{{\tt 1404.4091}}.

\bibitem{Casali:2014xpa}
E.~Casali, ``{Soft sub-leading divergences in Yang-Mills amplitudes},'' {\em
  JHEP} {\bf 08} (2014) 077,
\href{http://www.arXiv.org/abs/1404.5551}{{\tt 1404.5551}}.

\bibitem{Schwab:2014xua}
B.~U.~W. Schwab and A.~Volovich, ``{Subleading Soft Theorem in Arbitrary
  Dimensions from Scattering Equations},'' {\em Phys. Rev. Lett.} {\bf 113}
  (2014), no.~10, 101601,
\href{http://www.arXiv.org/abs/1404.7749}{{\tt 1404.7749}}.

\bibitem{Bern:2014oka}
Z.~Bern, S.~Davies, and J.~Nohle, ``{On Loop Corrections to Subleading Soft
  Behavior of Gluons and Gravitons},'' {\em Phys. Rev.} {\bf D90} (2014),
  no.~8, 085015,
\href{http://www.arXiv.org/abs/1405.1015}{{\tt 1405.1015}}.

\bibitem{He:2014bga}
S.~He, Y.-t. Huang, and C.~Wen, ``{Loop Corrections to Soft Theorems in Gauge
  Theories and Gravity},'' {\em JHEP} {\bf 12} (2014) 115,
\href{http://www.arXiv.org/abs/1405.1410}{{\tt 1405.1410}}.

\bibitem{Larkoski:2014hta}
A.~J. Larkoski, ``{Conformal Invariance of the Subleading Soft Theorem in Gauge
  Theory},'' {\em Phys. Rev.} {\bf D90} (2014), no.~8, 087701,
\href{http://www.arXiv.org/abs/1405.2346}{{\tt 1405.2346}}.

\bibitem{Cachazo:2014dia}
F.~Cachazo and E.~Y. Yuan, ``{Are Soft Theorems Renormalized?},''
\href{http://www.arXiv.org/abs/1405.3413}{{\tt 1405.3413}}.

\bibitem{Afkhami-Jeddi:2014fia}
N.~Afkhami-Jeddi, ``{Soft Graviton Theorem in Arbitrary Dimensions},''
\href{http://www.arXiv.org/abs/1405.3533}{{\tt 1405.3533}}.

\bibitem{Adamo:2014yya}
T.~Adamo, E.~Casali, and D.~Skinner, ``{Perturbative gravity at null
  infinity},'' {\em Class. Quant. Grav.} {\bf 31} (2014), no.~22, 225008,
\href{http://www.arXiv.org/abs/1405.5122}{{\tt 1405.5122}}.

\bibitem{Bianchi:2014gla}
M.~Bianchi, S.~He, Y.-t. Huang, and C.~Wen, ``{More on Soft Theorems: Trees,
  Loops and Strings},'' {\em Phys. Rev.} {\bf D92} (2015), no.~6, 065022,
\href{http://www.arXiv.org/abs/1406.5155}{{\tt 1406.5155}}.

\bibitem{Bern:2014vva}
Z.~Bern, S.~Davies, P.~Di~Vecchia, and J.~Nohle, ``{Low-Energy Behavior of
  Gluons and Gravitons from Gauge Invariance},'' {\em Phys.Rev.} {\bf D90}
  (2014), no.~8, 084035,
\href{http://www.arXiv.org/abs/1406.6987}{{\tt 1406.6987}}.

\bibitem{Broedel:2014fsa}
J.~Broedel, M.~de~Leeuw, J.~Plefka, and M.~Rosso, ``{Constraining subleading
  soft gluon and graviton theorems},'' {\em Phys.Rev.} {\bf D90} (2014), no.~6,
  065024,
\href{http://www.arXiv.org/abs/1406.6574}{{\tt 1406.6574}}.

\bibitem{He:2014cra}
T.~He, P.~Mitra, A.~P. Porfyriadis, and A.~Strominger, ``{New Symmetries of
  Massless QED},'' {\em JHEP} {\bf 10} (2014) 112,
\href{http://www.arXiv.org/abs/1407.3789}{{\tt 1407.3789}}.

\bibitem{Zlotnikov:2014sva}
M.~Zlotnikov, ``{Sub-sub-leading soft-graviton theorem in arbitrary
  dimension},'' {\em JHEP} {\bf 10} (2014) 148,
\href{http://www.arXiv.org/abs/1407.5936}{{\tt 1407.5936}}.

\bibitem{Kalousios:2014uva}
C.~Kalousios and F.~Rojas, ``{Next to subleading soft-graviton theorem in
  arbitrary dimensions},'' {\em JHEP} {\bf 01} (2015) 107,
\href{http://www.arXiv.org/abs/1407.5982}{{\tt 1407.5982}}.

\bibitem{Du:2014eca}
Y.-J. Du, B.~Feng, C.-H. Fu, and Y.~Wang, ``{Note on Soft Graviton theorem by
  KLT Relation},'' {\em JHEP} {\bf 11} (2014) 090,
\href{http://www.arXiv.org/abs/1408.4179}{{\tt 1408.4179}}.

\bibitem{Luo:2014wea}
H.~Luo, P.~Mastrolia, and W.~J. Torres~Bobadilla, ``{Subleading soft behavior
  of QCD amplitudes},'' {\em Phys. Rev.} {\bf D91} (2015), no.~6, 065018,
\href{http://www.arXiv.org/abs/1411.1669}{{\tt 1411.1669}}.

\bibitem{White:2014qia}
C.~White, ``{Diagrammatic insights into next-to-soft corrections},'' {\em
  Phys.Lett.} {\bf B737} (2014) 216--222,
\href{http://www.arXiv.org/abs/1406.7184}{{\tt 1406.7184}}.

\bibitem{Low:1958sn}
F.~E. Low, ``{Bremsstrahlung of very low-energy quanta in elementary particle
  collisions},'' {\em Phys. Rev.} {\bf 110} (1958)
974--977.

\bibitem{Burnett:1967km}
T.~H. Burnett and N.~M. Kroll, ``{Extension of the low soft photon theorem},''
  {\em Phys. Rev. Lett.} {\bf 20} (1968)
86.

\bibitem{DelDuca:1990gz}
V.~Del~Duca, ``High-energy bremsstrahlung theorems for soft photons,'' {\em
  Nucl. Phys.} {\bf B345} (1990)
369--388.

\bibitem{Laenen:2008gt}
E.~Laenen, G.~Stavenga, and C.~D. White, ``{Path integral approach to eikonal
  and next-to-eikonal exponentiation},'' {\em JHEP} {\bf 0903} (2009) 054,
\href{http://www.arXiv.org/abs/0811.2067}{{\tt 0811.2067}}.

\bibitem{Laenen:2010uz}
E.~Laenen, L.~Magnea, G.~Stavenga, and C.~D. White, ``{Next-to-eikonal
  corrections to soft gluon radiation: a diagrammatic approach},'' {\em JHEP}
  {\bf 1101} (2011) 141,
\href{http://www.arXiv.org/abs/1010.1860}{{\tt 1010.1860}}.

\bibitem{Bonocore:2015esa}
D.~Bonocore, E.~Laenen, L.~Magnea, S.~Melville, L.~Vernazza, and C.~D. White,
  ``{A factorization approach to next-to-leading-power threshold logarithms},''
  {\em JHEP} {\bf 06} (2015) 008,
\href{http://www.arXiv.org/abs/1503.05156}{{\tt 1503.05156}}.

\bibitem{Bonocore:2014wua}
D.~Bonocore, E.~Laenen, L.~Magnea, L.~Vernazza, and C.~D. White, ``{The method
  of regions and next-to-soft corrections in Drell-Yan production},'' {\em
  Phys.Lett.} {\bf B742} (2015) 375--382,
\href{http://www.arXiv.org/abs/1410.6406}{{\tt 1410.6406}}.

\bibitem{Akhoury:2013yua}
R.~Akhoury, R.~Saotome, and G.~Sterman, ``{High Energy Scattering in
  Perturbative Quantum Gravity at Next to Leading Power},''
\href{http://www.arXiv.org/abs/1308.5204}{{\tt 1308.5204}}.

\bibitem{Bjerrum-Bohr:2016hpa}
N.~E.~J. Bjerrum-Bohr, J.~F. Donoghue, B.~R. Holstein, L.~Plante, and
  P.~Vanhove, ``{Light-like Scattering in Quantum Gravity},''
\href{http://www.arXiv.org/abs/1609.07477}{{\tt 1609.07477}}.

\bibitem{Gardi:2009zv}
E.~Gardi and L.~Magnea, ``{Infrared singularities in QCD amplitudes},'' {\em
  Nuovo Cim.} {\bf 032C} (2009) 137--157,
\href{http://www.arXiv.org/abs/0908.3273}{{\tt 0908.3273}}.

\bibitem{Catani:1996jh}
S.~Catani and M.~Seymour, ``{The Dipole formalism for the calculation of QCD
  jet cross-sections at next-to-leading order},'' {\em Phys.Lett.} {\bf B378}
  (1996) 287--301,
\href{http://www.arXiv.org/abs/hep-ph/9602277}{{\tt hep-ph/9602277}}.

\bibitem{Catani:1996vz}
S.~Catani and M.~Seymour, ``{A General algorithm for calculating jet
  cross-sections in NLO QCD},'' {\em Nucl.Phys.} {\bf B485} (1997) 291--419,
\href{http://www.arXiv.org/abs/hep-ph/9605323}{{\tt hep-ph/9605323}}.

\bibitem{Brezin:1970zr}
E.~Brezin, C.~Itzykson, and J.~Zinn-Justin, ``{Relativistic balmer formula
  including recoil effects},'' {\em Phys.Rev.} {\bf D1} (1970)
2349--2355.

\bibitem{Kabat:1992tb}
D.~N. Kabat and M.~Ortiz, ``{Eikonal quantum gravity and Planckian
  scattering},'' {\em Nucl.Phys.} {\bf B388} (1992) 570--592,
\href{http://www.arXiv.org/abs/hep-th/9203082}{{\tt hep-th/9203082}}.

\bibitem{Weinberg:1965nx}
S.~Weinberg, ``{Infrared photons and gravitons},'' {\em Phys.Rev.} {\bf 140}
  (1965)
B516--B524.

\bibitem{Akhoury:2011kq}
R.~Akhoury, R.~Saotome, and G.~Sterman, ``{Collinear and Soft Divergences in
  Perturbative Quantum Gravity},'' {\em Phys.Rev.} {\bf D84} (2011) 104040,
\href{http://www.arXiv.org/abs/1109.0270}{{\tt 1109.0270}}.

\bibitem{Beneke:2012xa}
M.~Beneke and G.~Kirilin, ``{Soft-collinear gravity},'' {\em JHEP} {\bf 1209}
  (2012) 066,
\href{http://www.arXiv.org/abs/1207.4926}{{\tt 1207.4926}}.

\bibitem{Gellas:1998sh}
G.~C. Gellas, A.~I. Karanikas, and C.~N. Ktorides, ``{Worldline approach to
  eikonals for QED and linearized quantum gravity and their off mass shell
  extensions},'' {\em Phys. Rev.} {\bf D57} (1998)
3763--3776.

\bibitem{Gross:1968in}
D.~J. Gross and R.~Jackiw, ``{Low-Energy Theorem for Graviton Scattering},''
  {\em Phys. Rev.} {\bf 166} (1968)
1287--1292.

\bibitem{DeWitt:1967uc}
B.~S. DeWitt, ``{Quantum Theory of Gravity. 3. Applications of the Covariant
  Theory},'' {\em Phys. Rev.} {\bf 162} (1967)
1239--1256.

\bibitem{Bjerrum-Bohr:2013bxa}
N.~E.~J. Bjerrum-Bohr, J.~F. Donoghue, and P.~Vanhove, ``{On-shell Techniques
  and Universal Results in Quantum Gravity},'' {\em JHEP} {\bf 02} (2014) 111,
\href{http://www.arXiv.org/abs/1309.0804}{{\tt 1309.0804}}.

\bibitem{BjerrumBohr:2004mz}
N.~E. Bjerrum-Bohr, {\em {Quantum gravity, effective fields and string
  theory}}.
\newblock PhD thesis, Bohr Inst., 2004.
\newblock
\href{http://www.arXiv.org/abs/hep-th/0410097}{{\tt hep-th/0410097}}.
\newblock

\bibitem{Bern:2013yya}
Z.~Bern, S.~Davies, T.~Dennen, Y.-t. Huang, and J.~Nohle, ``{Color-Kinematics
  Duality for Pure Yang-Mills and Gravity at One and Two Loops},'' {\em Phys.
  Rev.} {\bf D92} (2015), no.~4, 045041,
\href{http://www.arXiv.org/abs/1303.6605}{{\tt 1303.6605}}.

\bibitem{Aichelburg:1970dh}
P.~C. Aichelburg and R.~U. Sexl, ``{On the Gravitational field of a massless
  particle},'' {\em Gen. Rel. Grav.} {\bf 2} (1971)
303--312.

\bibitem{Veneziano:2009zz}
G.~Veneziano, ``{Transplanckian string collisions: An update},'' in {\em {On
  recent developments in theoretical and experimental general relativity,
  astrophysics and relativistic field theories. Proceedings, 12th Marcel
  Grossmann Meeting on General Relativity, Paris, France, July 12-18, 2009.
  Vol. 1-3}}, pp.~95--107.
\newblock
2009.
\newblock

\end{thebibliography}\endgroup
\end{document}